\documentclass[journal]{IEEEtran}
\usepackage{cite}
\usepackage{amsmath,amssymb,amsfonts}
\usepackage{amsmath}
\usepackage{mathrsfs}
\usepackage{graphicx}
\usepackage{textcomp}
\usepackage[table,xcdraw]{xcolor}
\usepackage{ragged2e}
\usepackage{bbm}
\usepackage{multirow}
\usepackage{hyperref}
\usepackage[linesnumbered, ruled]{algorithm2e}
\usepackage{algpseudocode}
\usepackage[caption=false, font=footnotesize]{subfig}
\usepackage{mathtools} 
\usepackage[hang,flushmargin]{footmisc}
\usepackage{lipsum}

\makeatletter
\def\ps@IEEEtitlepagestyle{%
  \def\@oddfoot{\mycopyrightnotice}%
  \def\@oddhead{\hbox{}\@IEEEheaderstyle\leftmark\hfil\thepage}\relax
  \def\@evenhead{\@IEEEheaderstyle\thepage\hfil\leftmark\hbox{}}\relax
  \def\@evenfoot{}%
}
\def\mycopyrightnotice{%
  \begin{minipage}{\textwidth}
  \scriptsize 1530-437X \textcopyright 2020 IEEE. Personal use of this material is permitted.
  Permission from IEEE must be obtained for all other uses, in any current or future
  media, including reprinting/republishing this material for advertising or promotional
  purposes, creating new collective works, for resale or redistribution to servers or
  lists, or reuse of any copyrighted component of this work in other works.
  DOI: \href{https://doi.org/10.1109/JSEN.2020.3015781}{10.1109/JSEN.2020.3015781}
  \end{minipage}
}
\makeatother

\makeatletter
\newcommand{\algorithmfootnote}[2][\footnotesize]{%
  \let\old@algocf@finish\@algocf@finish
  \def\@algocf@finish{\old@algocf@finish
    \leavevmode\rlap{\begin{minipage}{\linewidth}
    #1#2
    \end{minipage}}%
  }%
}
\makeatother
\usepackage[british]{babel}
\usepackage[autostyle]{csquotes}
\usepackage{placeins}

\usepackage{float}

\def\BibTeX{{\rm B\kern-.05em{\sc i\kern-.025em b}\kern-.08em
    T\kern-.1667em\lower.7ex\hbox{E}\kern-.125emX}}

\begin{document}

\title{A Novel Multi-Stage Training Approach for Human Activity Recognition from Multimodal Wearable Sensor Data Using Deep Neural Network}

\author{Tanvir~Mahmud,~\IEEEmembership{Student~Member,~IEEE,}
        A.~Q.~M.~Sazzad~Sayyed,~\IEEEmembership{Student~Member,~IEEE,}
        Shaikh~Anowarul~Fattah,~\IEEEmembership{Senior~Member,~IEEE,}
        and~Sun-Yuan~Kung,~\IEEEmembership{Life~Fellow,~IEEE}
\thanks{T.~Mahmud, A.~Q.~M.~S.~Sayyed and S.~A.~Fattah are with the Department of Electrical and Electronic Engineering, BUET, Dhaka 1000,
Bangladesh (e-mail: tanvirmahmud@eee.buet.ac.bd; fattah@eee.buet.ac.bd).}
\thanks{S.~Y.~Kung is with the Department of Electrical Engineering, Princeton University, USA (e-mail: kung@princeton.edu)}
}

\markboth{IEEE Sensors Journal,~Vol.~21, No.~2, January~2021}%
{Tanvir \MakeLowercase{\textit{et al.}}: A Novel Multi-Stage Training Approach for Human Activity Recognition from Multimodal Wearable Sensor Data Using Deep Neural Network}

\maketitle
\begin{abstract}
    Deep neural network is an effective choice to automatically recognize human actions utilizing data from various wearable sensors. These networks automate the process of feature extraction relying completely on data. However, various noises in time series data with complex inter-modal relationships among sensors make this process more complicated. In this paper, we have proposed a novel multi-stage training approach that increases diversity in this feature extraction process to make accurate recognition of actions by combining varieties of features extracted from diverse perspectives. Initially, instead of using single type of transformation, numerous transformations are employed on time series data to obtain variegated representations of the features encoded in raw data. An efficient deep CNN architecture is proposed that can be individually trained to extract features from different transformed spaces. Later, these CNN feature extractors are merged into an optimal architecture finely tuned for optimizing diversified extracted features through a combined training stage or multiple sequential training stages. This approach offers the opportunity to explore the encoded features in raw sensor data utilizing multifarious observation windows with immense scope for efficient selection of features for final convergence. Extensive experimentations have been carried out in three publicly available datasets that provide outstanding performance consistently with average five-fold cross-validation accuracy of $\mathbf{99.29\%}$ on UCI HAR database, $\mathbf{99.02\%}$ on USC HAR database, and $\mathbf{97.21\%}$ on SKODA database outperforming other state-of-the-art approaches.
\end{abstract}


\IEEEpeerreviewmaketitle

\section{Introduction}
\IEEEPARstart{A}{ctivity} recognition using wearable sensors has been a trending topic of research for its widespread applicability on diverse domains ranging from health care services to military applications~\cite{a1}. With the ubiquitous availability of modern mobile devices such as smartphones, tablets, and smartwatches, various types of sensor data are available that can be utilized effectively in numerous applications like activity recognition.    
Various types of sensor data along with image and video data have been employed for recognizing human activity~\cite{i1}. In this work, we have mainly focused on the time series wearable sensor data, e.g. accelerometer, gyroscope, and magnetometer, as these are easy to obtain even with our smart devices and can be used to recognize human activity from distant position on real-time basis as these sensors' data are very small in volume and easy to share through internet.

\begin{figure*}[!t]
    \centering
    \subfloat[]{\includegraphics[scale=0.38]{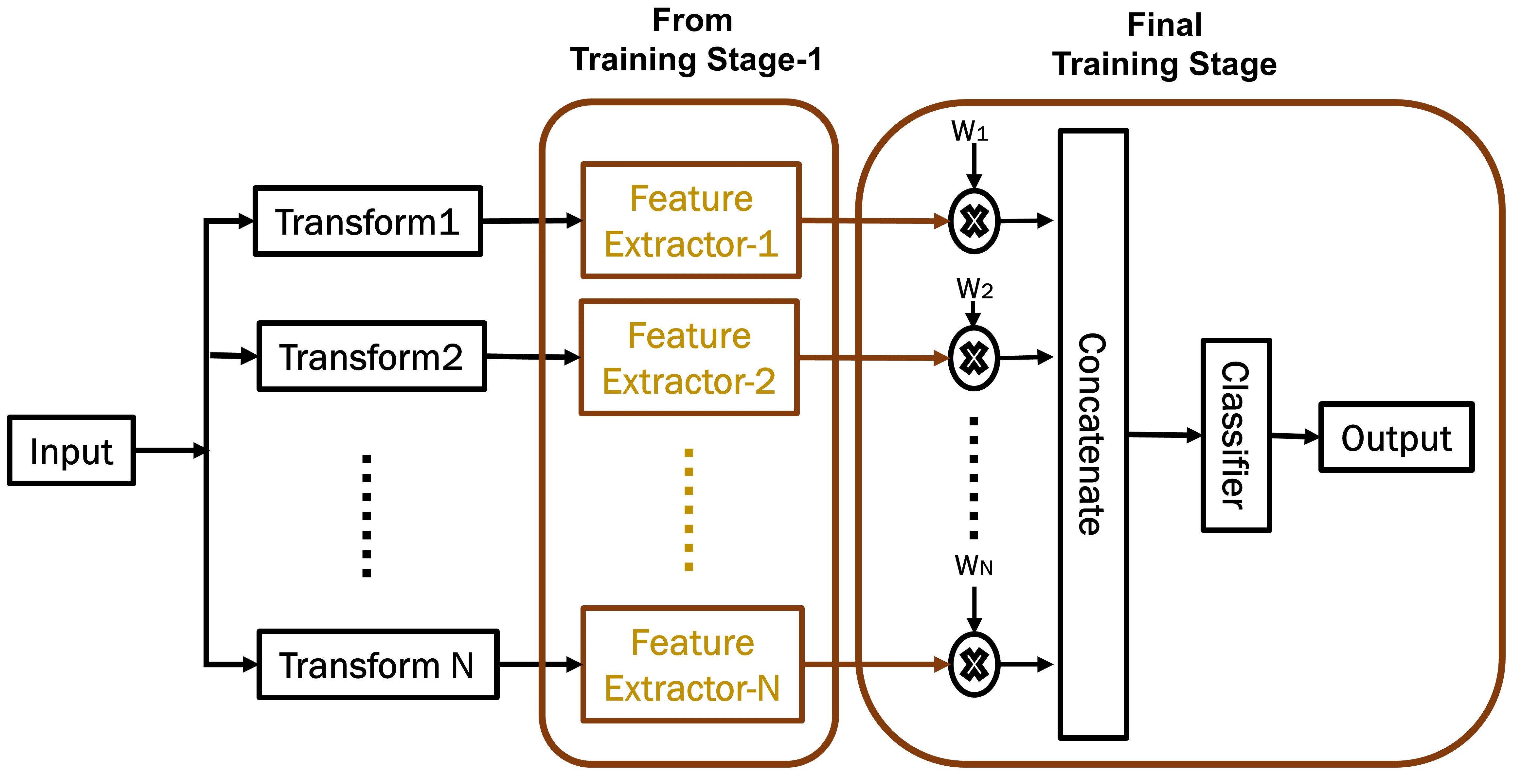}%
    \label{bb}}
    \hspace{0.3cm}
    \subfloat[]{\includegraphics[scale=0.35]{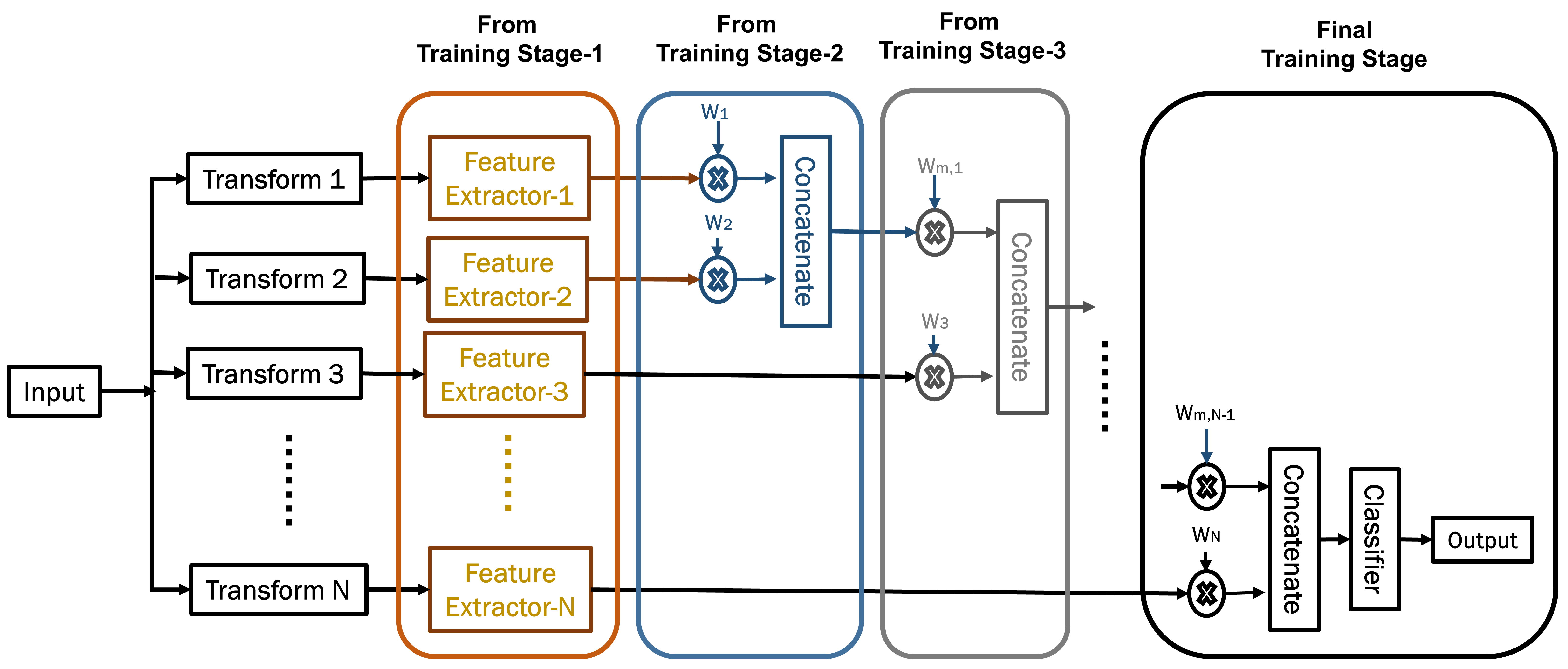}%
    \label{bc}}
    \caption{\textbf{Multiple training stages are utilized to incorporate features from numerous transformed representations of input sensor data. In training stage-1, different feature extractors are individually trained to extract features from different transformed spaces. Features extracted from diverse representation spaces can be weighted and these weight vectors can be jointly optimized either (a) in a combined additional training stage, or (b) multiple sequential training stages can be utilized to learn the weight vectors in a sequential manner.}}    
    \label{f1}
\end{figure*}

Large varieties of approaches have been applied to make the correct recognition ranging from traditional feature-based
approaches to the end-to-end deep neural network in recent times. Numerous hand-crafted feature extraction process with shallow classifiers are explored in the literature for utilizing multimodal sensor data in activity recognition~\cite{r6,n2,m29,m23,n3,m31,m6,m30}. 
Though these types of handcrafted features perform well in limited training data scenario, the extraction of effective features gets very complicated with more number of sensors. Additionally, the process heavily demands domain expertise for proper selection of features which becomes harder with the presence of random noises that occurs very often in practical conditions.

To automate the complicated feature extraction process, various types of deep neural networks have been studied in the literature to recognize human activity from wearable sensor data~\cite{m28,rr1,r8, m11,n6,n5,n1, r2, r9, r5,r7}. Most of these approaches directly employ the collected raw sensor data for automated feature extraction using the deep neural network, such as convolutional neural network (CNN)~\cite{m28,rr1, r8}, recurrent neural network (RNN)~\cite{n1,r2}, long short term memory (LSTM) network~\cite{r9}, hybrid CNN-LSTM network~\cite{r5}, and a more complicated LSTM-CNN-LSTM based hierarchical network~\cite{r7}. 
Most of these networks are very deep in structure and therefore, a large amount of data is required to train them properly. Moreover, due to random noises and perturbations in multi-modal sensor data from different sources, the process gets more intricate to operate with the raw data directly.  
Hence, with an increasing number of sensors, while having a small amount of data for some of the activity classes, this problem becomes critical for the automated extraction of features from raw sensor data using deep network that severely affects the performance.

In~\cite{g1,r1,w2,s}, various approaches have been introduced to represent the time series data in a modified space that makes the feature extraction process easier by reducing the effects of noise or random variations. These transformations on the time series sensor data provide more opportunities to explore the variations of features from different spaces. 
Though these transformations provide efficient representation of some of the features in a different space, some other features may become complicated to extract from that particular space. However, different transformations provide diverse viewpoints to explore the feature space of raw time series data. Hence, similar to these studies, depending solely on a single transformed space for feature exploration limits the scope of feature extraction that may result in smaller performance in many circumstances. If features extracted from different transformed spaces can be incorporated in the final decision-making process, it will provide a more robust opportunity to analyze the information on raw data. But, the challenging task of integrating effective features from diverse transformed spaces through joint optimization to reach the optimum performance in activity recognition is yet to be attempted.

In this work, we have proposed a novel multi-stage training methodology to overcome these limitations by efficiently employing a multitude of time-series transformations that facilitates the exploration of diversified feature spaces.
Instead of relying on a single transformed space, features from numerous transformed spaces are integrated together. An efficient deep convolutional neural network architecture is proposed that can be separately tuned and optimized as efficient feature extractors from different transformed spaces. These CNN based automated feature extractors reduce the complexity of manual feature selection from numerous transformed spaces. Afterwards, these separately optimized networks operating on different transformed spaces are combined into a unified architecture utilizing the proposed additional combined training stage or multi-stage sequential training stages where features extracted from different transformed spaces are sequentially weighted, optimized, and integrated for the final activity recognition.  Hence, different portions of this unified architecture are trained and optimized in several training stages that make it possible to optimize with a smaller amount of available training data.  Moreover, different types of realistic data augmentation techniques have been introduced to increase the variations of the available data. The proposed approach opens scopes for optimization of diversified features extracted from different transformed spaces and makes the process more resilient from noise and other random perturbations by exploiting the advantages provided by numerous representations of the raw data. 
Results of intense experimentations have been presented using three publicly available datasets that provide very satisfactory performance compared to other state-of-the-art approaches.

 \renewcommand{\theenumi}{\roman{enumi}}

\section{Methodology}
The proposed multi-stage training approach is represented in Fig.~\ref{f1}. In the first stage of training, individual feature extractors operating on different transformed spaces are trained in parallel with separate classifiers. In the literature, varieties of feature extractors from time-series data have been explored ranging from PCA, ICA, wavelet-based methods to modern CNN, DNN, LSTM, and numerous deep learning methods~\cite{m28,n1,n5,n6}. To overcome additional complexities mainly arising from the difficulty of feature selection and optimization from different diversified transformed representations of time series data, we have proposed deep CNN architectures as feature extractors from different transformed domains.  As it is completely data-driven, deep CNN architecture can be trained as an efficient feature extractor from any representation of data. For joint optimization of multiple transformed feature spaces, learning of this first training stage is transferred into a unified structure utilizing another combined training stage (Fig.~\ref{bb}) or utilizing a number of sequential training stages (Fig.~\ref{bc}).

After completing the first stage of training, all the separate classifiers of individual networks are removed. As a result, when input data is fed to the network, representational features extracted from different transformed domains utilizing the trained feature extractors are available which were fed into separate classifier units in the first training stage. However, the feature quality can be varied with the transformation of the raw data which can be visible by evaluating the performance of the separate feature extractors in the first stage. Hence, in the second and final stage of training (Fig.~\ref{bb}), these feature vectors are multiplied by separate weighting vectors to increase the selectivity of the system.  Later, all these weighted feature vectors are concatenated together and a common dense classifier unit is trained to provide the exact prediction from these concatenated features. Therefore, these weighting vectors, along with the combined dense classifier unit, are supposed to learn in this stage of training utilizing the data again.      
\begin{figure}[!t]
    \centering
    \subfloat[]{\includegraphics[scale=0.43]{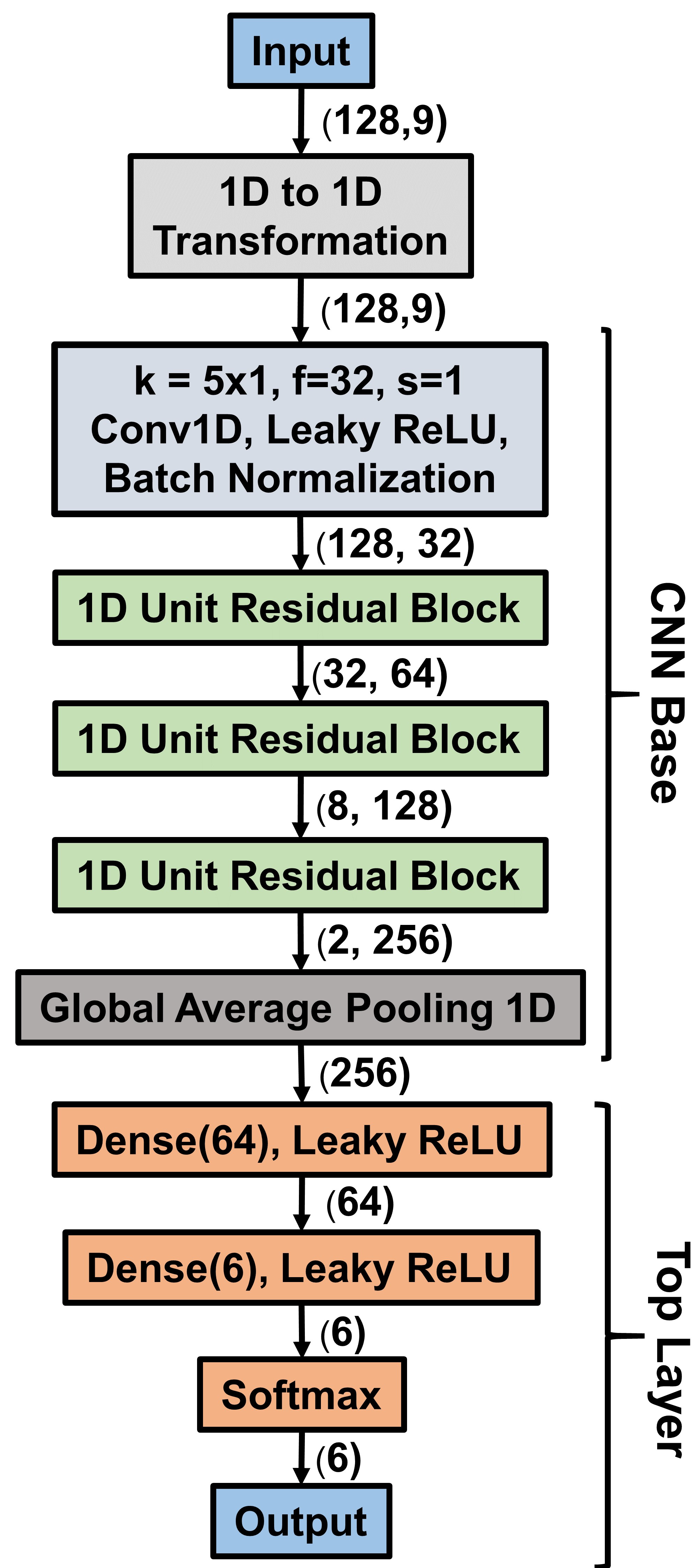}%
    \label{da}}
    \hspace{0.7cm}
    \subfloat[]{\includegraphics[scale=0.44]{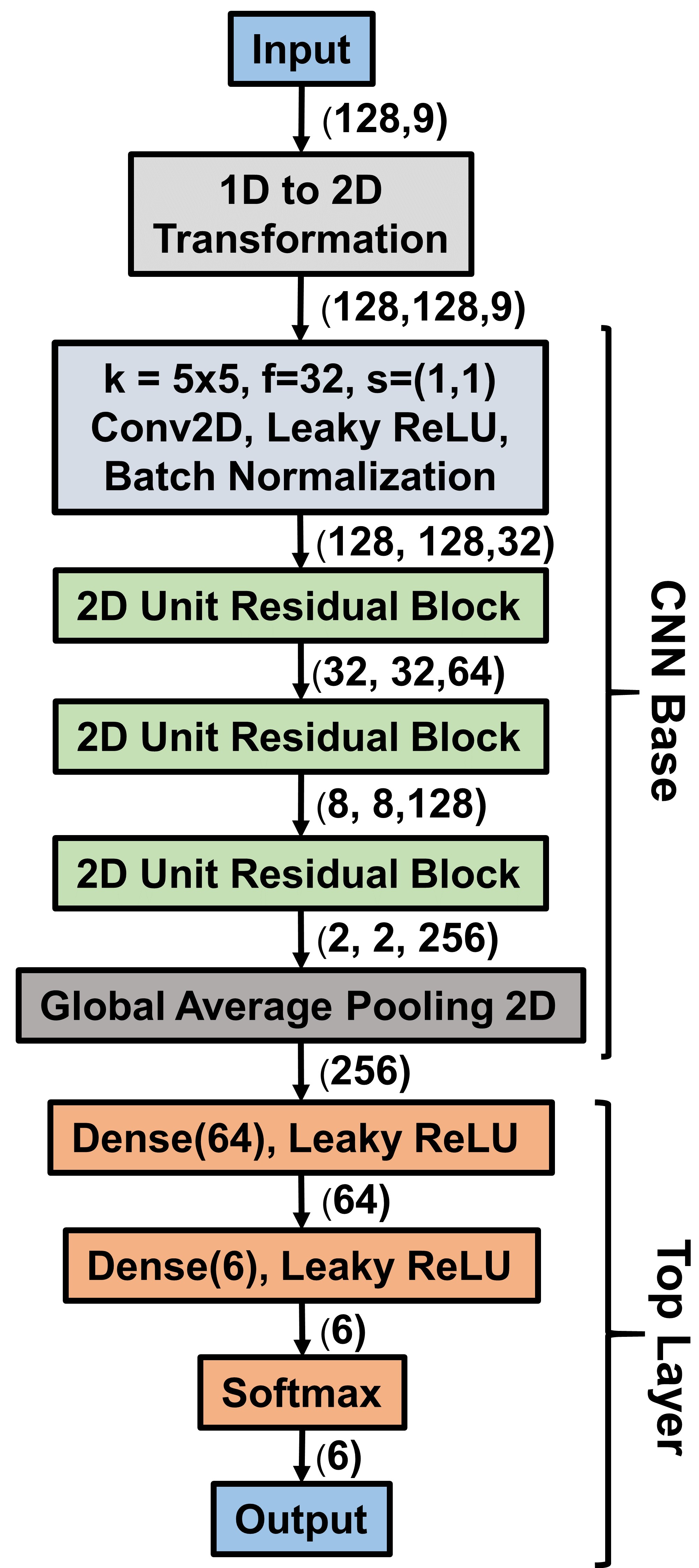}%
    \label{db}}
    \caption{\textbf{Proposed (a) $\mathbf{1D}$ Convolutional Neural Network and (b) $\mathbf{2D}$ Convolutional Neural Network. Tensor dimensions shown after each operation are optimized for UCI HAR Database~\cite{m19}.}}
    \label{f3}
\end{figure}

\begin{figure}[!t]
    \centering
    \subfloat[]{\includegraphics[scale=0.4]{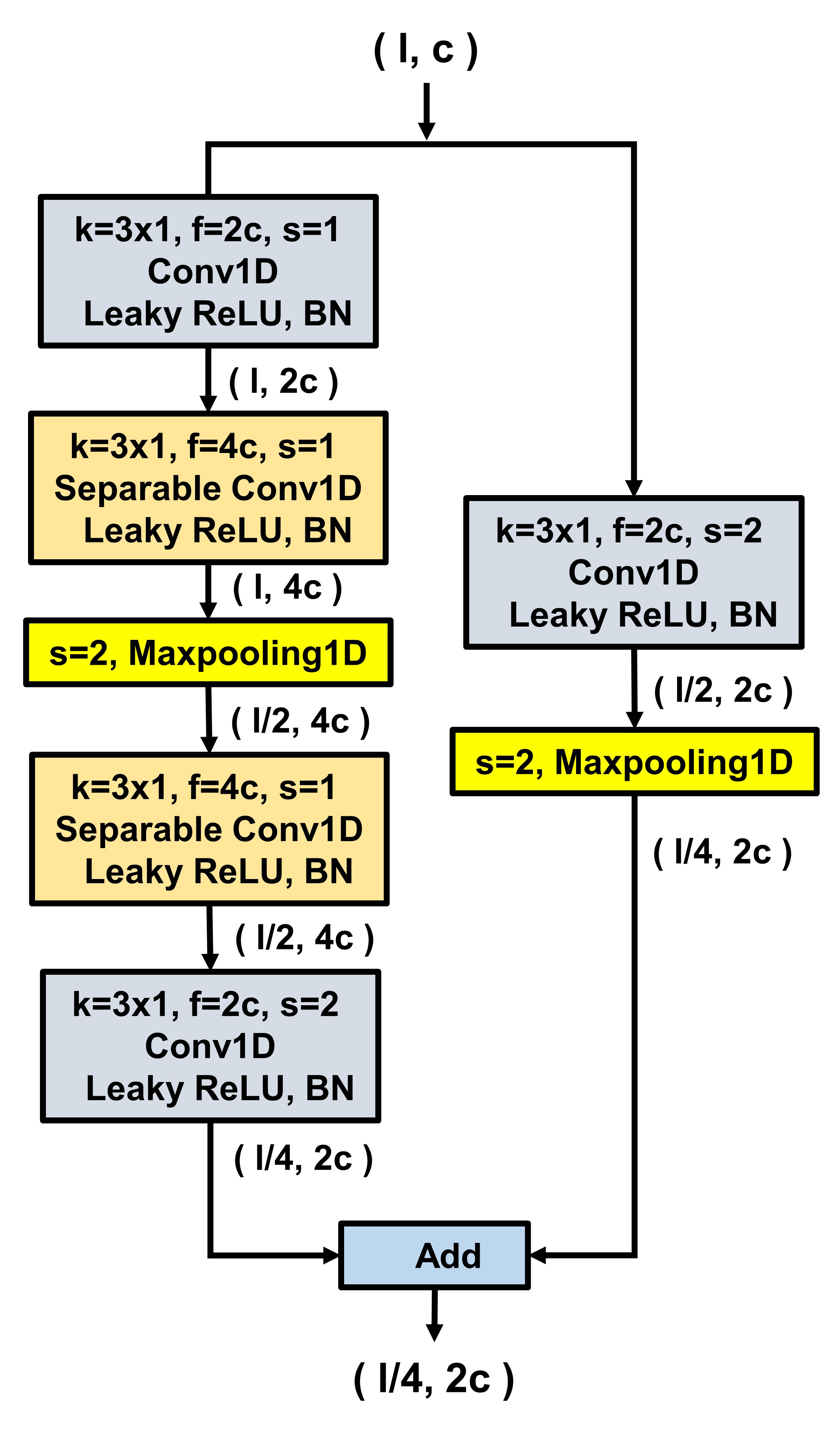}%
    \label{ca}}
    \hfill
    \subfloat[]{\includegraphics[scale=0.4]{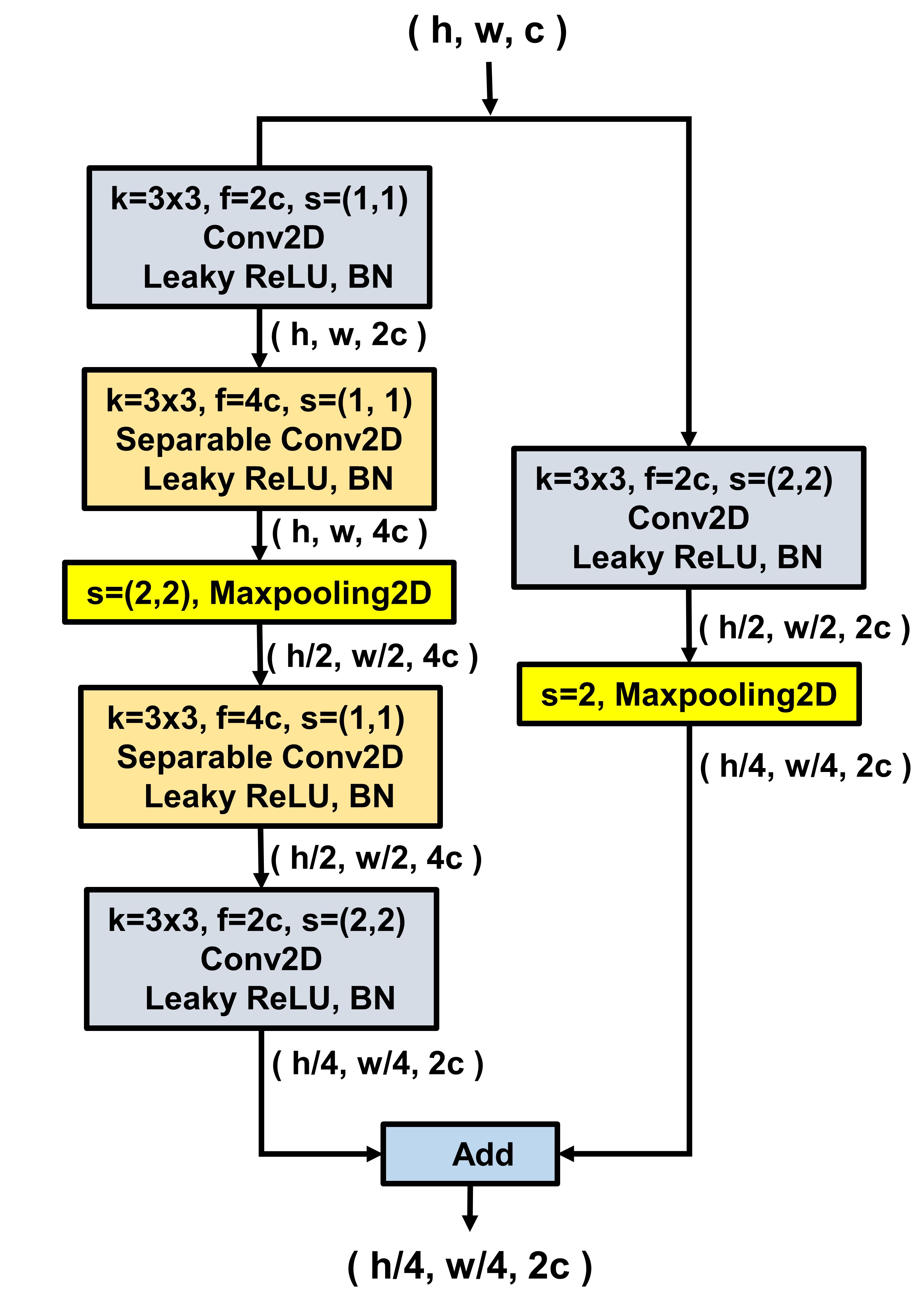}%
    \label{cb}}
    \caption{\textbf{Proposed (a) $\mathbf{1D}$ unit residual block and (b) $\mathbf{2D}$ unit residual block. In (a), `l' denotes length and `c' denotes number of channels of the $\mathbf{1D}$ tensor. While in (b), `h' and `w' denote height and width of the $\mathbf{2D}$ tensor, respectively. Additionally, `k' stands for kernel size, `f' for filter number, `s' for strides, and `BN' for batch normalization in the convolution.}}
    \label{f2}
\end{figure}
In Fig.~\ref{bc}, the proposed multi-stage sequential training is shown. In the two-stage training, as described, the final second-stage training learns the weighting vector for each feature map at the same time with the combined classifier. However, in the multi-stage sequential training,
weighting vectors for only two feature vectors, extracted by the feature extractors trained in the previous stage, are learned along with the combined classifier at a time. In the following stage, the classifier is removed and the merged weighted feature vectors of these two transformations undergo through similar next stage of training with one of the remaining feature vectors representing different transformation. Thus, in each stage of sequential training, weighted feature vector from an additional transformed space is accumulated with the combined feature extractors trained in the previous stage. This method of sequential training offers additional opportunity to converge individual feature representations corresponding to variegated transformed spaces to the final decision-making process by optimizing two feature vectors sequentially. Moreover, in deep learning-based approaches, these weighting vectors applied on separate feature vectors can be easily integrated by introducing a separate densely connected layer operating on each feature vectors accompanied by different weighting vectors. 

\subsection{Transformations on Time Series Data}
Different types of transformations on time series data have been utilized in the proposed approach. These are described briefly as below.
\subsubsection{Gramian Angular Field Transformation (GAF)}
Gramian angular field transformation maps the elements of a $1D$ time-series data into a $2D$ matrix representation. This encoding scheme preserves the temporal dependency of the original time series data along the diagonal of the encoded matrix while the non-diagonal entries essentially represent the correlation between samples~\cite{g1}. In this transformation, $G:R^N  \rightarrow R^{N \times N}$, the input time series, $X$, is transformed into polar coordinate $(r,\phi)$ after normalization. 
\begin{align}
    \centering
    \phi_{i} &= \cos^{-1}({x_{i}}),\ -1\leq {x_{i}}\leq 1,\  {x}_i\ \epsilon\ {X} \\
    r_{i} &= \frac{t_{i}}{N},\ t_{i}\ \epsilon\  \mathbb{N}
\end{align}
Here, $t_{i}$ the time stamp
and $N$ is a constant factor to regularize the span of the polar coordinate system. These polar angles  are utilized to get the final transformed matrix $G$, which is,
\begin{equation}
    G_{i,j} = cos(\phi_{i}+\phi_{j}), \ i,j= 1,2,\dots, n
\end{equation}
    
\begin{figure*}[!t]
    \centering
    \subfloat[]{\includegraphics[scale=0.32]{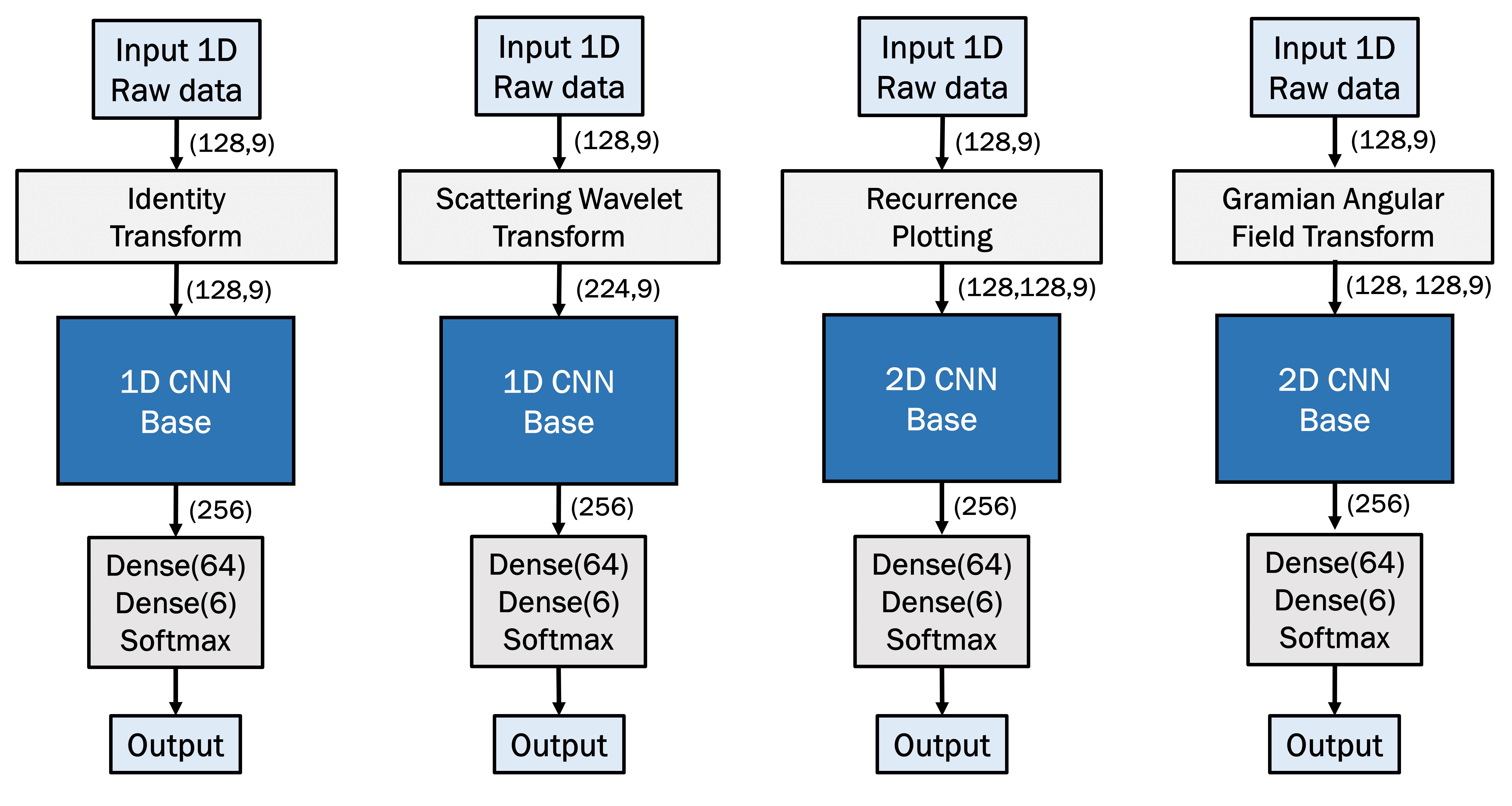}%
    \label{t1a}}
    \hspace{.5cm}
    \subfloat[]{\includegraphics[scale=0.3]{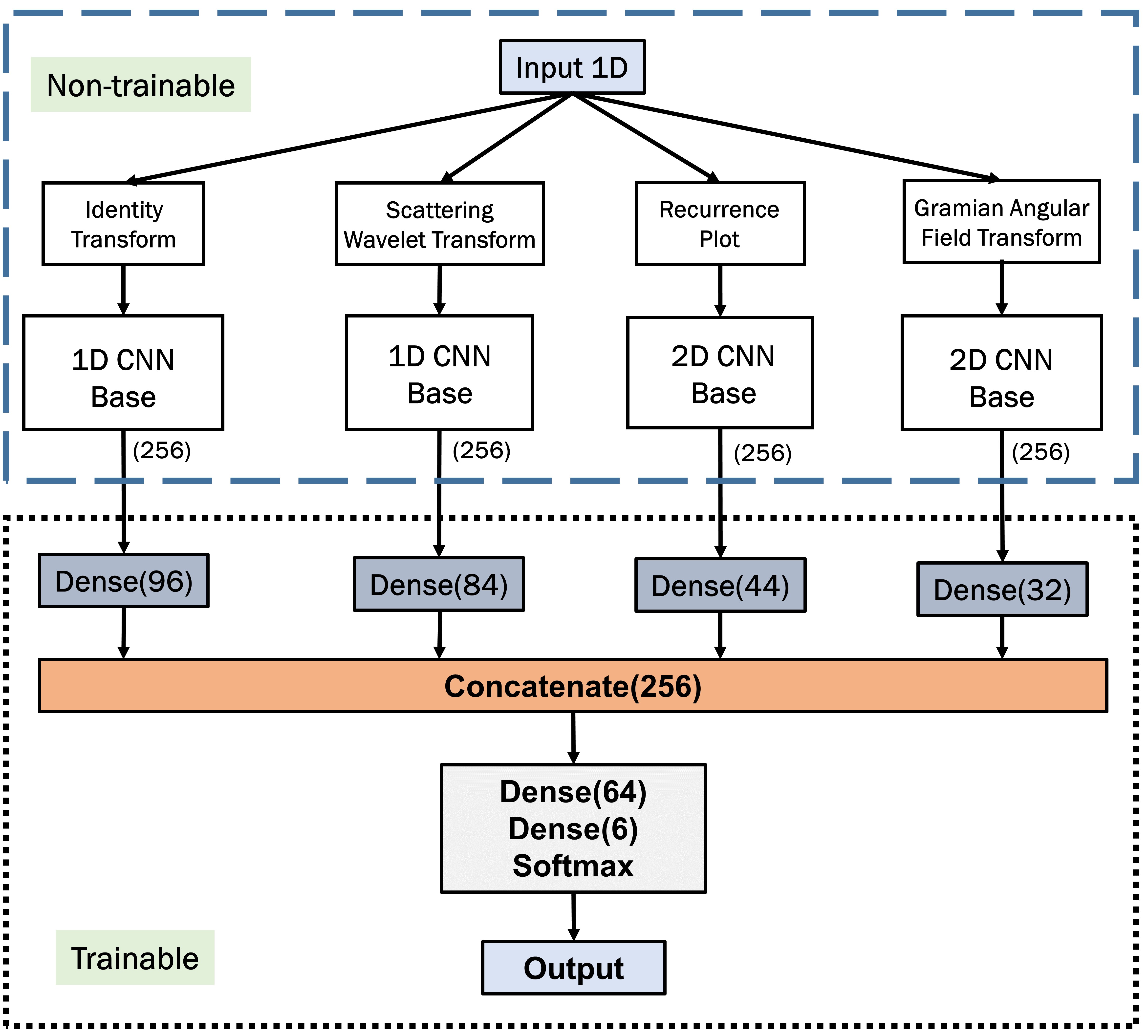}%
    \label{t1b}}
    \vfill
    \subfloat[]{\includegraphics[scale=0.305]{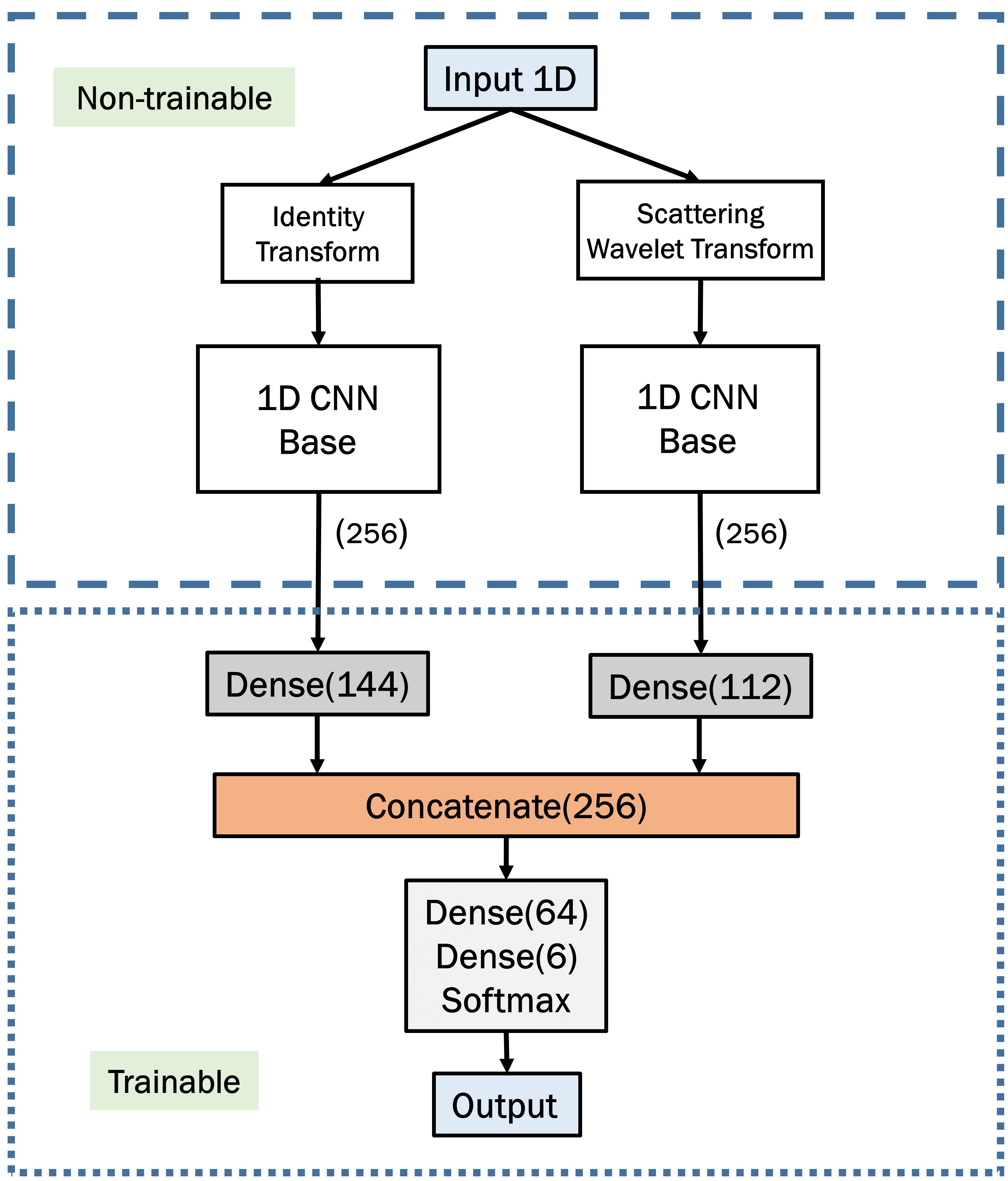}%
    \label{t2b}}
    \hspace{0.3cm}
    \subfloat[]{\includegraphics[scale=0.295]{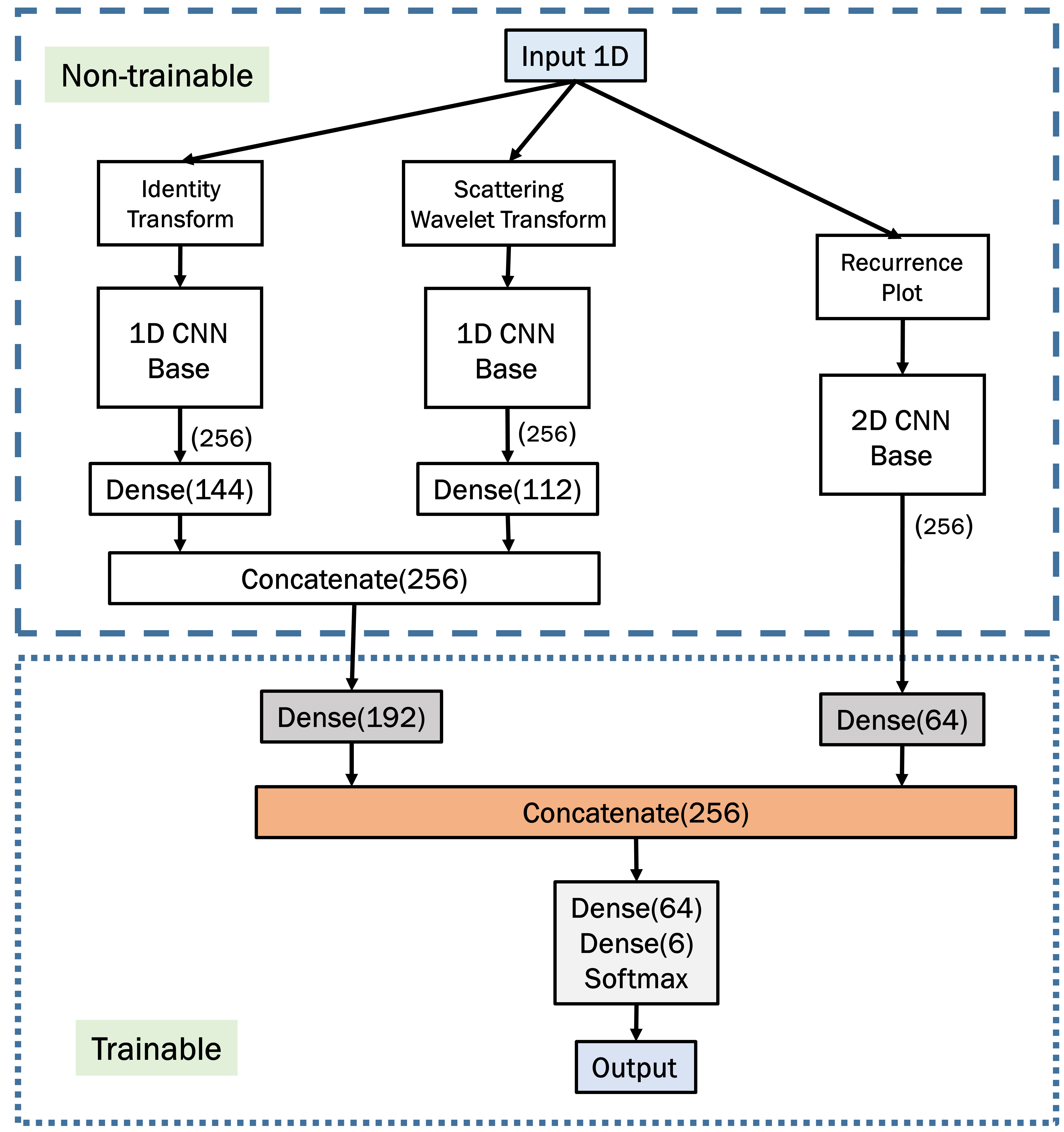}%
    \label{t2c}}
    \hspace{0.3cm}
    \subfloat[]{\includegraphics[scale=0.29]{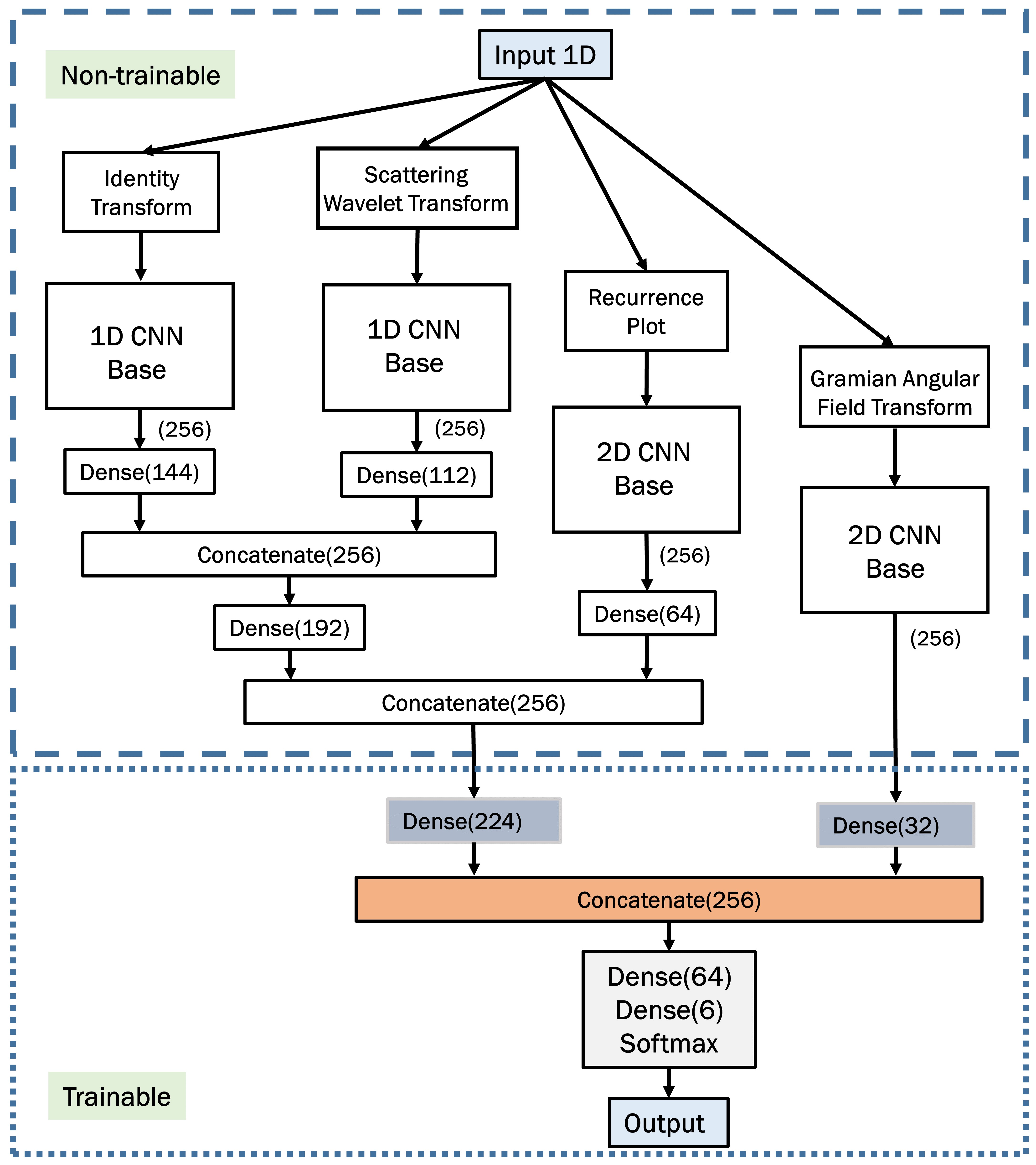}%
    \label{t2d}}
    \caption{\textbf{Schematic representation of the proposed multi-stage sequential training scheme.
    Here, (a) represents Individual training stage, (b) represents combined training stage where all the pre-trained networks are merged using one additional training stage, and (c), (d), (e) represent the sequential training stages where pre-trained networks are converged sequentially towards the unified architecture. The tensor dimensions are optimized for UCI HAR dataset~\cite{m19}.}}
    \label{t2}
\end{figure*}
    
\subsubsection{Recurrence Plotting}
The recurrence plot portrays the inherent recurrent behavior of time-series, e.g. irregular cyclicity and periodicity, into a $2D$ matrix~\cite{r1}. This method provides a way to explore the m-dimensional phase space trajectory of time series data for generating a $2D$ representation by searching points of some trajectories that have returned to the previous state and is represented by,
\begin{equation}
    \centering
     R_{i,j} = \theta(\epsilon - || \mathbf{s_i- s_j} ||),\ \mathbf{s}(.) \epsilon\ R^{m},\ i,j = 1,2,\dots,K
\end{equation}

where $K$ is the number of considered states $\mathbf{s}$, $\epsilon$ is a threshold distance, $||.||$ a norm and $\theta(.)$ is the Heaviside
function.

\subsubsection{Scattering Wavelet Transformation}
Scattering wavelet transform offers representational features of the time-series data those are rotation/translation-invariant while remaining stable to deformations. This technique provides the opportunity to extract features from a very small number of data ~\cite{w2}. A mortlet wavelet function, defined as mother wavelet, undergoes through convolution operation with the raw time series data while being scaled and rotated, and thus creates different levels of representational features. 

Let's consider, $W_j$ and $U_j$ to be the averaging operation and complex modulus of the averaged signal, respectively, for order $j$ ($0,1,\dots,L$) of the scattering coefficients, and these coefficients can be described as
\begin{equation}
    \centering
    S_j = W_j U_j S_{j-1}* \lvert \psi _j \rvert*\phi _j,
\end{equation}
where $\phi_j$ represents the Gaussian low pass filter and $\psi_j$ represents the mortlet wavelet function of order $j$.
Therefore, a scattering representation, $S_X$ of  time series data, $X$, is obtained by concatenating the scattering coefficients of a different order,
\begin{equation}
    S_X= [S_0X, S_1X, \dots , S_LX]    
\end{equation}

As multi-channel sensor data collected from numerous sensors have been used in this work, each channel of such time-series data is transformed individually using any of these transformations, and all such transformed data are stacked together maintaining a similar time information in all the channels. Later, they undergo through the feature extraction process utilizing deep neural networks. 

\begin{algorithm}[!t]
\KwData{training sample $\emph{\textbf{X}}$; training label $y_{actual}$}
\KwResult{weight matrices $\textbf{\emph{D}}, \textbf{\emph{F}}$}
\tcc{Individual training begins}
\For{$i\leftarrow 1$ \KwTo $N$}{
    \ Calculate $\hat{X_i} = T_i(X)$\;
    \ Randomize $D_{1,i}^l$ and $F_{i}$ , for $l=[1,\dots,L]$\;
    \While{The training error threshold is unsatisfied}{
        \ Calculate $f_i = F_{i}(\hat{X_{i}})$\;
        \ Find $ y_{pred,i}^1 = D_{1,i}^L(D_{1,i}^{L-1}(\dots(D_{1,i}^1(f_i))))$\;
        \ Find loss $L_{1,i} = \mathscr{L}(y_{pred,i}^1, y_{actual})$\;
        \ Update $D_{1,i}^l$ and $F_{i}$, for $l=[1,\dots,L]$ \;
    }
    \ Calculate $d_i = F_{i}(\hat{X_{i}}) $\;
}
\tcc{Combined training stage begins}
\ Randomize $D_2^m$, for $m=1,\dots, L' $\;
\While{The training error threshold is unsatisfied}{
    \For {$i\leftarrow 1$ \KwTo $N$}{
        \ Set, $f_{i} = D_{2,i}^1(d_i)$\;
    }
    \ Set feature mapping group, $ f = [f_1, f_2, \dots, f_N] $\;
    \ Find $y_{pred}^{2} = D_{2}^{L'}(D_{2}^{L'-1}(\dots(D_{2}^2(f))))$\;
    \ Find loss $L_2 = \mathscr{L}(y_{pred}^2, y_{actual})$\;
    \ Update ${\emph{D}_2^m}$, for $m=1,\dots, L'$ \;
}
\caption{Proposed Two-Stage Training Method}
\algorithmfootnote{$F_i$ denotes the CNN base part of $i^{th}$ transform.\\
$D_{n}^{l}$ denotes the $l^{th}$ densely connected layer of $n_{th}$ training stage.\\
$T_i$ denotes the $i_{th}$ transformation on raw data.}
\label{a1}
\end{algorithm}

\begin{algorithm}[tb]
\KwData{training sample $\emph{\textbf{X}}$; training label $y_{actual}$}
\KwResult{weight matrices $\textbf{\emph{D}}, \textbf{\emph{F}}$}
\tcc{Individual training begins}
\For{$i\leftarrow 1$ \KwTo $N$}{
    \ Calculate $\hat{X_i} = T_i(X)$\;
    \ Randomize $D_{1,i}^l$ and $F_{i}$ , for $l=[1,\dots,L]$\;
    \While{The training error threshold is unsatisfied}{
        \ Calculate $f_i = F_{i}(\hat{X_{i}})$\;
        \ Find $ y_{pred,i}^1 = D_{1,i}^L(D_{1,i}^{L-1}(\dots(D_{1,i}^1(f_i))))$\;
        \ Find loss $L_1,i = \mathscr{L}(y_{pred,i}^1, y_{actual})$\;
        \ Update $D_{1,i}^l$ and $F_{i}$, for $l=[1,\dots,L]$ \;
    }
}
\tcc{Sequential training begins}
\ Initialize $F_{merged,1} = F_1$ \;
\For{$n\leftarrow 2$ \KwTo $N$}{
    \ Set $\hat{X}_{merged,n} = [\hat{X}_1, \dots, \hat{X}_{n-1}]$ \;
    \ Randomize $D_n^m$, for $m=1,\dots, L' $\;
    \While{The training error threshold is unsatisfied}{
        \ Set $f_{1,n} = D_{n}^1(F_{merged,n-1}(\hat{X}_{merged,n}))$\;
        \ Set $f_{2,n} = D_{n}^2(F_{n}(\hat{X}_n)) $ \;
        \ Set feature mapping group, $ f_n = [ f_{1,n}, f_{2,n} ]$\;
        \ Find $y_{pred}^{n} = D_{n}^{L'}(D_{n}^{L'-1}(\dots(D_{n}^3(f_n))))$\;
        \ Find loss $L_n = \mathscr{L}(y_{pred}^n, y_{actual})$\;
        \ Update weights of ${\emph{D}_n^m}$, for $m=1,\dots, L'$\;
    }
    \ Calculate $\hat{F}_{n-1}= D_{n}^1\circ F_{merged,n-1}$\;
    \ Calculate $\hat{F}_{n} = D_{n}^2\circ F_{n}$\;
    \ Set $F_{merged,n} = [\hat{F}_{n-1},\ \hat{F}_{n}]$\;
}
\caption{Proposed Sequential Training Method}
\algorithmfootnote{$F_i$ denotes the CNN base part for $i^{th}$ transform.\\
$D_{n}^{l}$ denotes the $l^{th}$ densely connected layer of $n_{th}$ training stage.\\
$T_i$ denotes the $i_{th}$ transformation on raw data. }
\label{a2}
\end{algorithm}

\subsection{Proposed Deep Neural Network Architectures}
 For feature extraction and classification, two deep CNN architectures are proposed, as shown in Fig. \ref{da} and Fig. \ref{db}, optimized to operate in 1D and 2D domain, respectively. Both of them are very similar to each other, as the objective of them is to extract features for activity recognition, with some modifications to operate in different domains for handling different dimensions of data. In general, the proposed CNN architecture mainly consists of a CNN base part followed by a top classifier layer. The CNN base part involves a number of convolution and pooling operations while the top classifier layer consists of a series of densely connected layers followed by the final activation layer to generate activity prediction. The operations performed here are discussed below.
 
 \begin{enumerate}
     \item The input 1D time-series data undergo an initial transformation operation as discussed above before starting the convolutional filtering in the deep network.
     
     \item Next, the tensor enters the convolutional base part where it passes through a series of unit residual block operations to extract deep features from a broad spectrum. Different representations of these unit residual blocks are shown in  Fig.~\ref{f2} with some variations in operations for handling $1D$ (Fig.~\ref{ca}) and $2D$ (Fig.~\ref{cb}) data. In these blocks, the input tensor passes through two different operations in parallel and the transformed tensors get added later to produce the final output tensor. Subsequently, a global average pooling operation is performed to extract the global features from each channel of the transformed tensor. This CNN base part extracts effective temporal/spatial features through convolutional filtering and pooling operations required for the final decision.  
     \item After that, the tensor propagates through the top classifier block where series of densely connected layers explore the extracted features of the CNN base part to get higher level of representation with the softmax activation layer at the end to merge these representations into a specified class of action. 
\end{enumerate}

The values of different convolutional kernel sizes, number of convolutional layers in each unit block, and number of unit residual blocks are established through experimentation to reach the maximum performance. Shallower networks are prone to underfit with the training data while deeper networks are prone to overfit. However, the proposed network effectively utilizes efficient separable convolutions along with residual operations to reduce vanishing gradient and overfitting issues for achieving optimum performance.

\begin{figure*}[!t]
    \centering
    \subfloat[]{\includegraphics[scale=0.15]{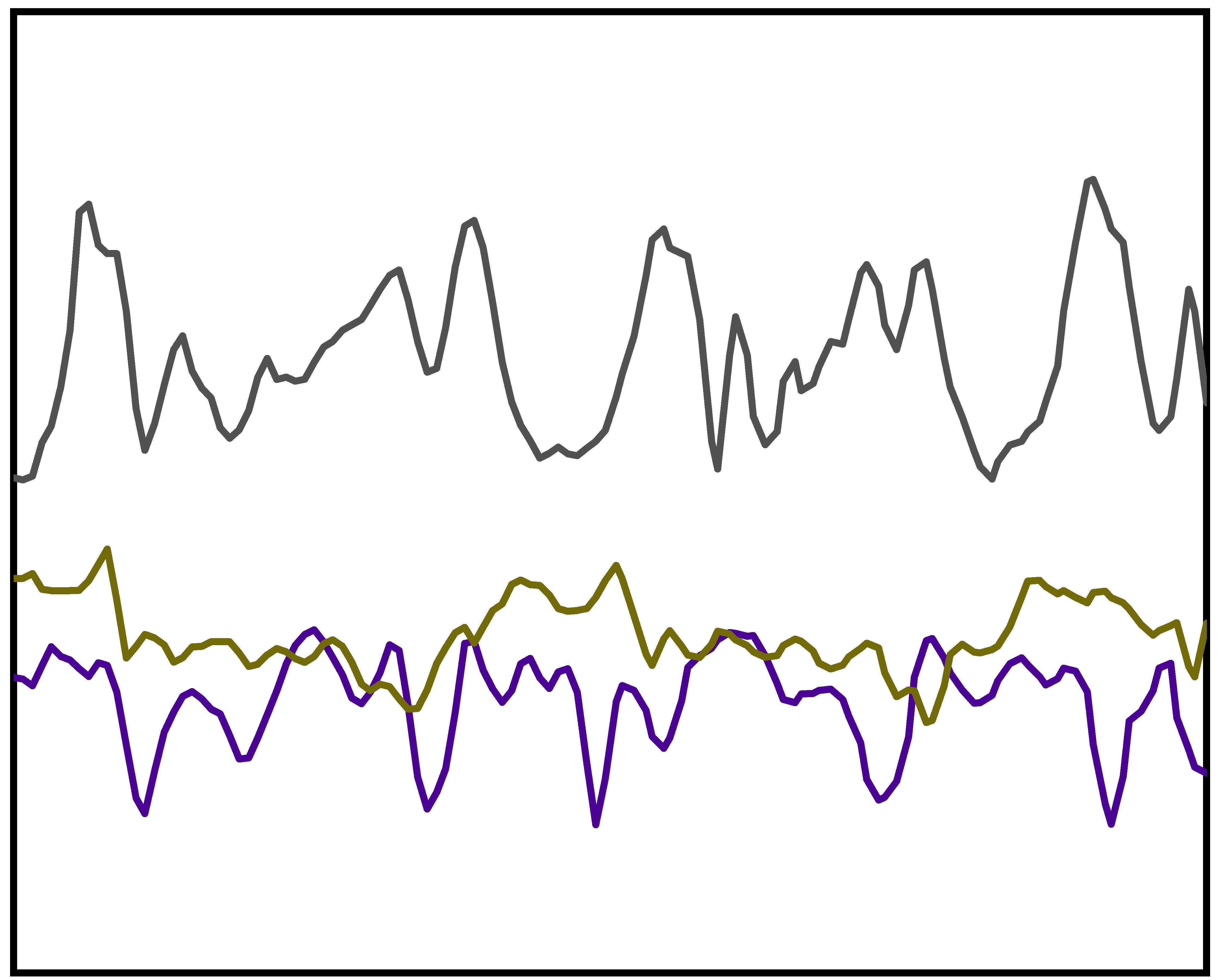}
    \label{a2a}}
    \hspace{1mm}
    \subfloat[]{\includegraphics[scale=0.15]{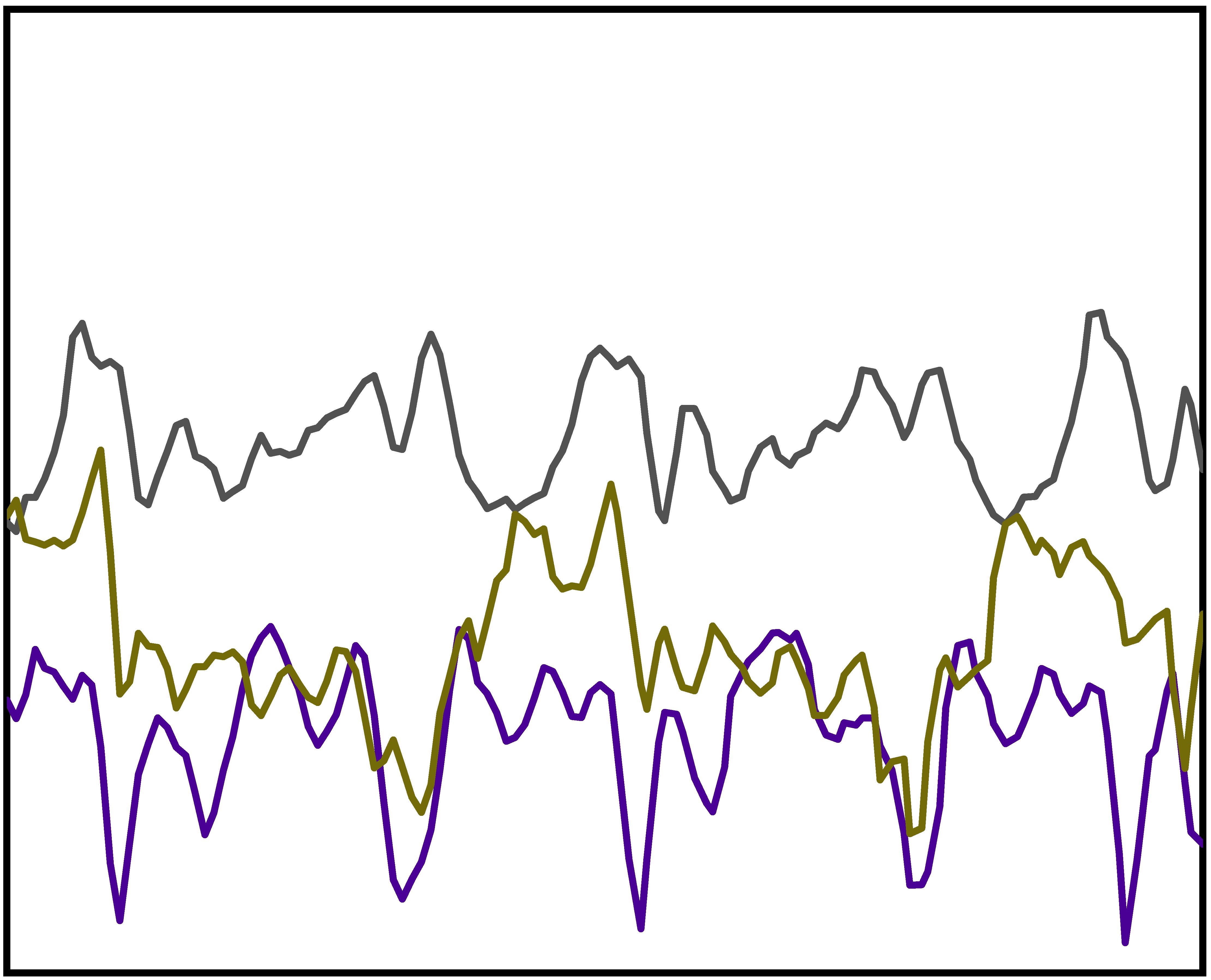}%
    \label{a2b}}
    \hspace{1mm}
    \subfloat[]{\includegraphics[scale=0.15]{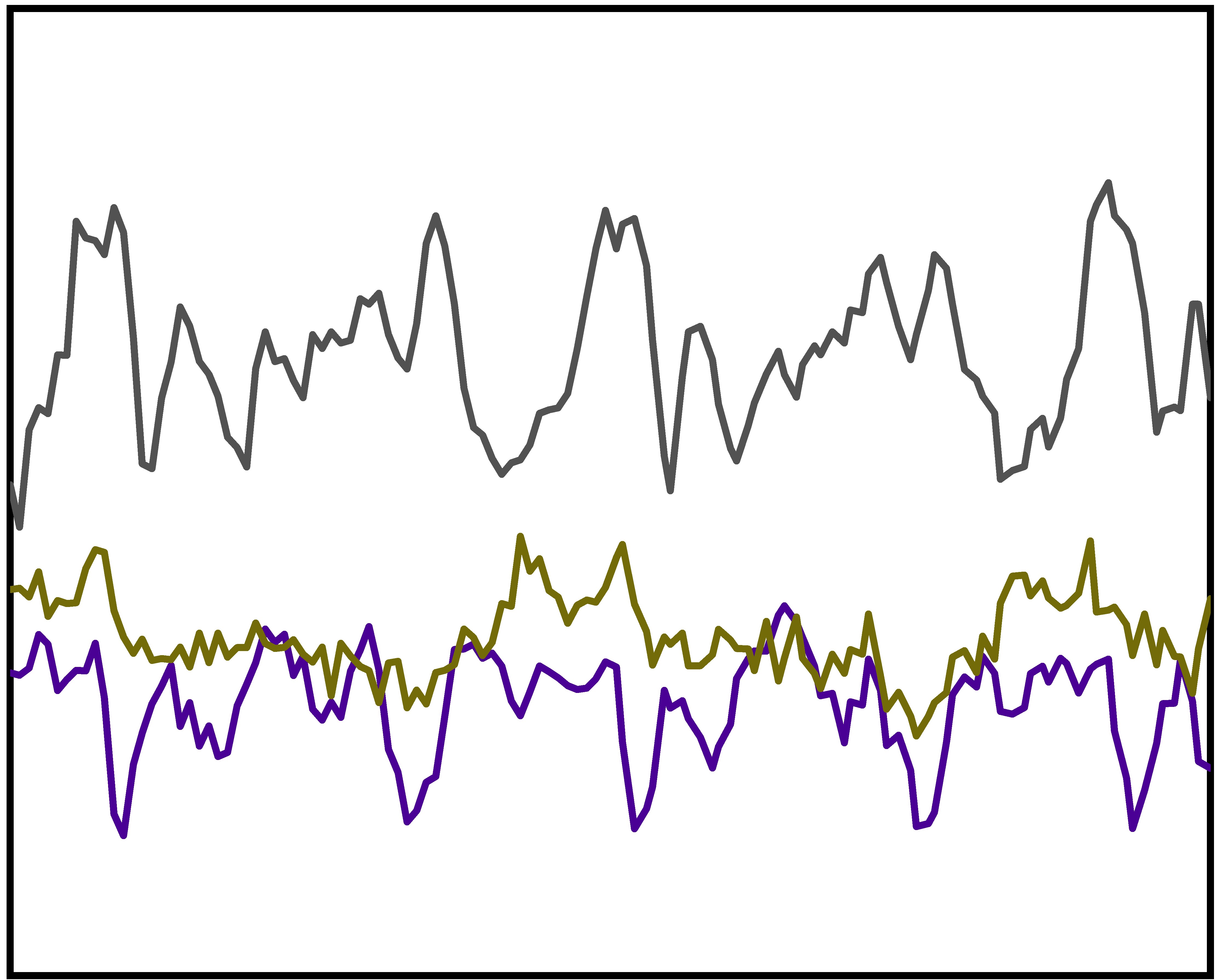}%
    \label{a2c}}
    \hspace{1mm}
    \subfloat[]{\includegraphics[scale=0.15]{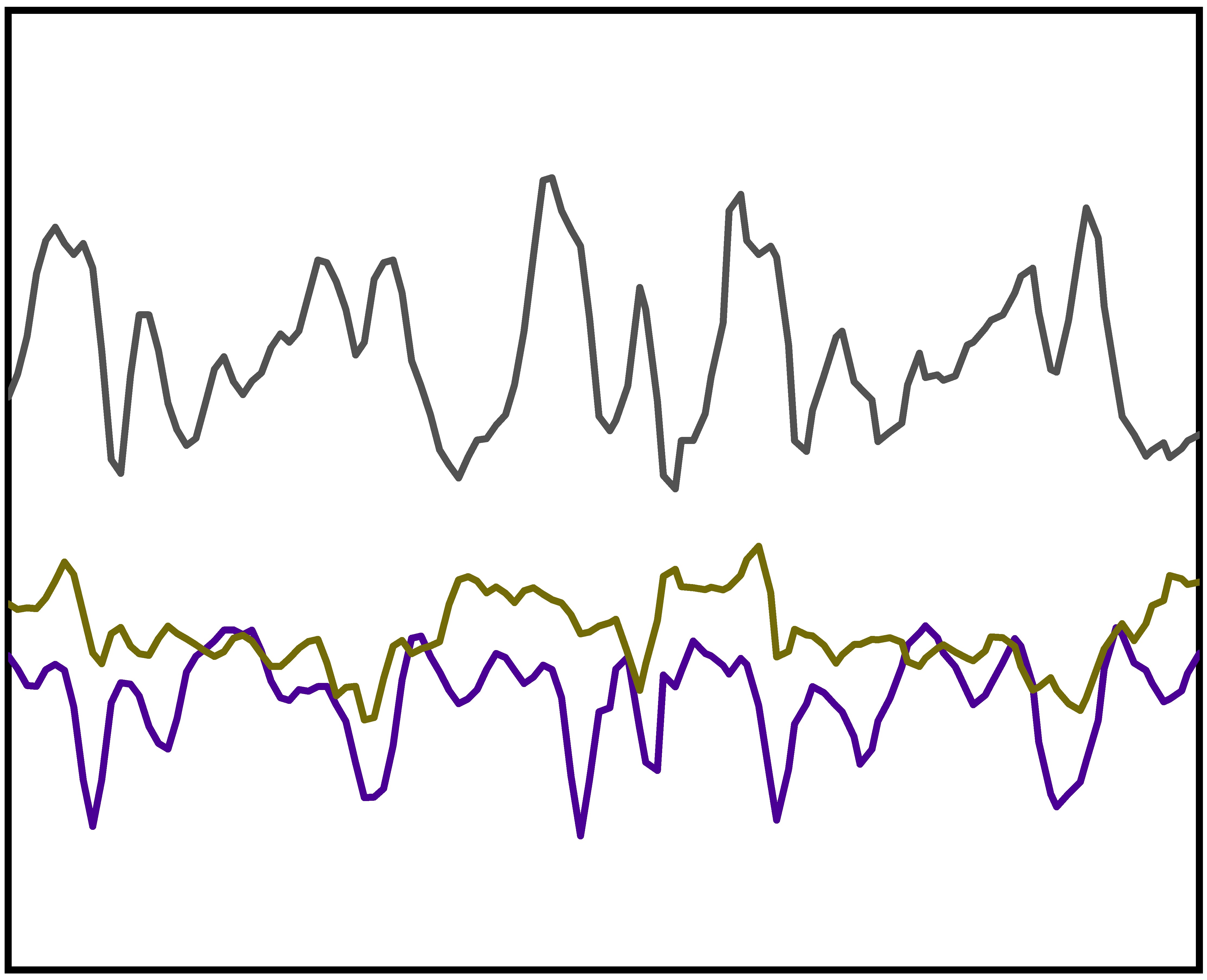}%
    \label{a2d}}
    \hspace{1mm}
    \subfloat[]{\includegraphics[scale=0.15]{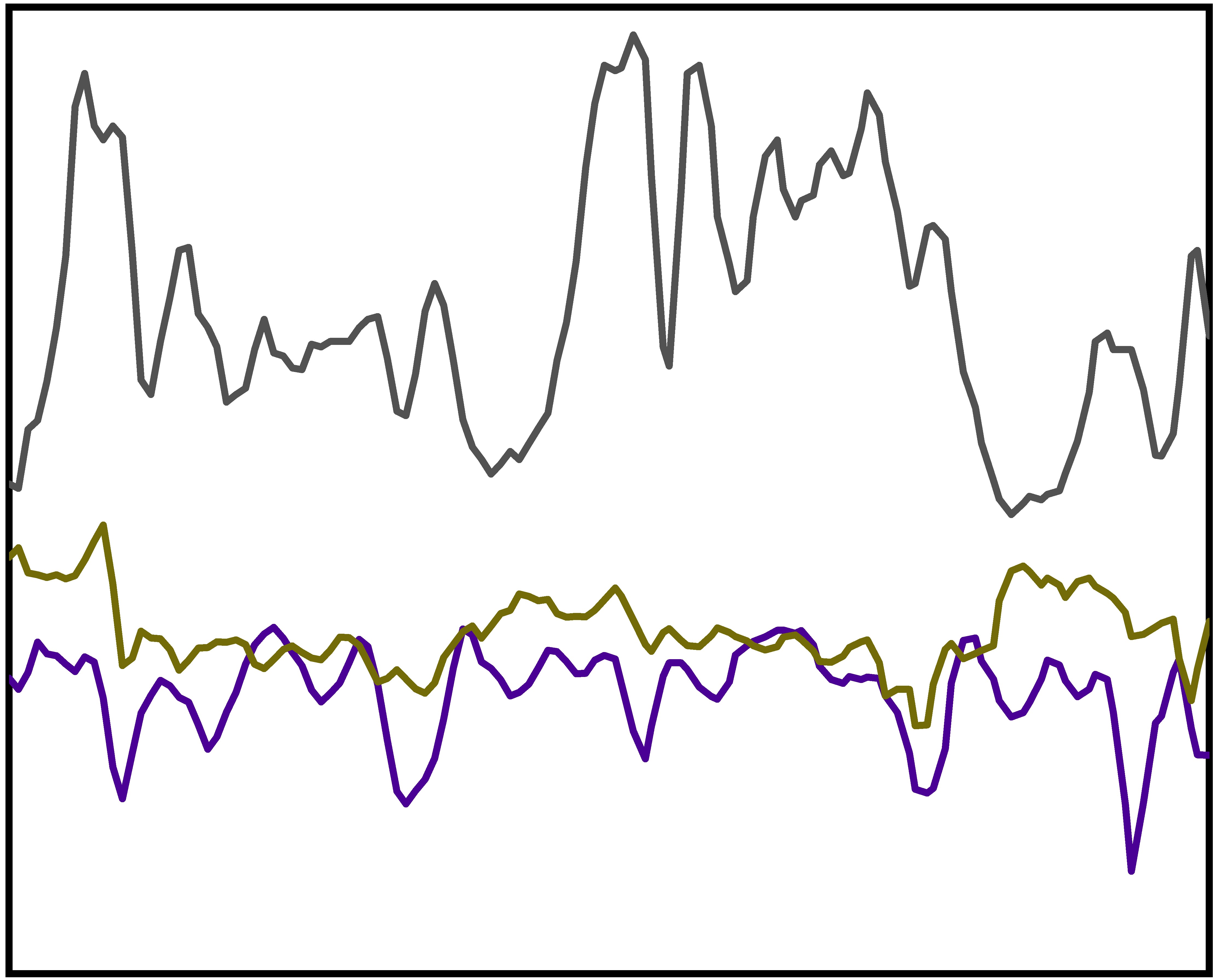}%
    \label{a2e}}
    \hspace{1mm}
    \subfloat[]{\includegraphics[scale=0.15]{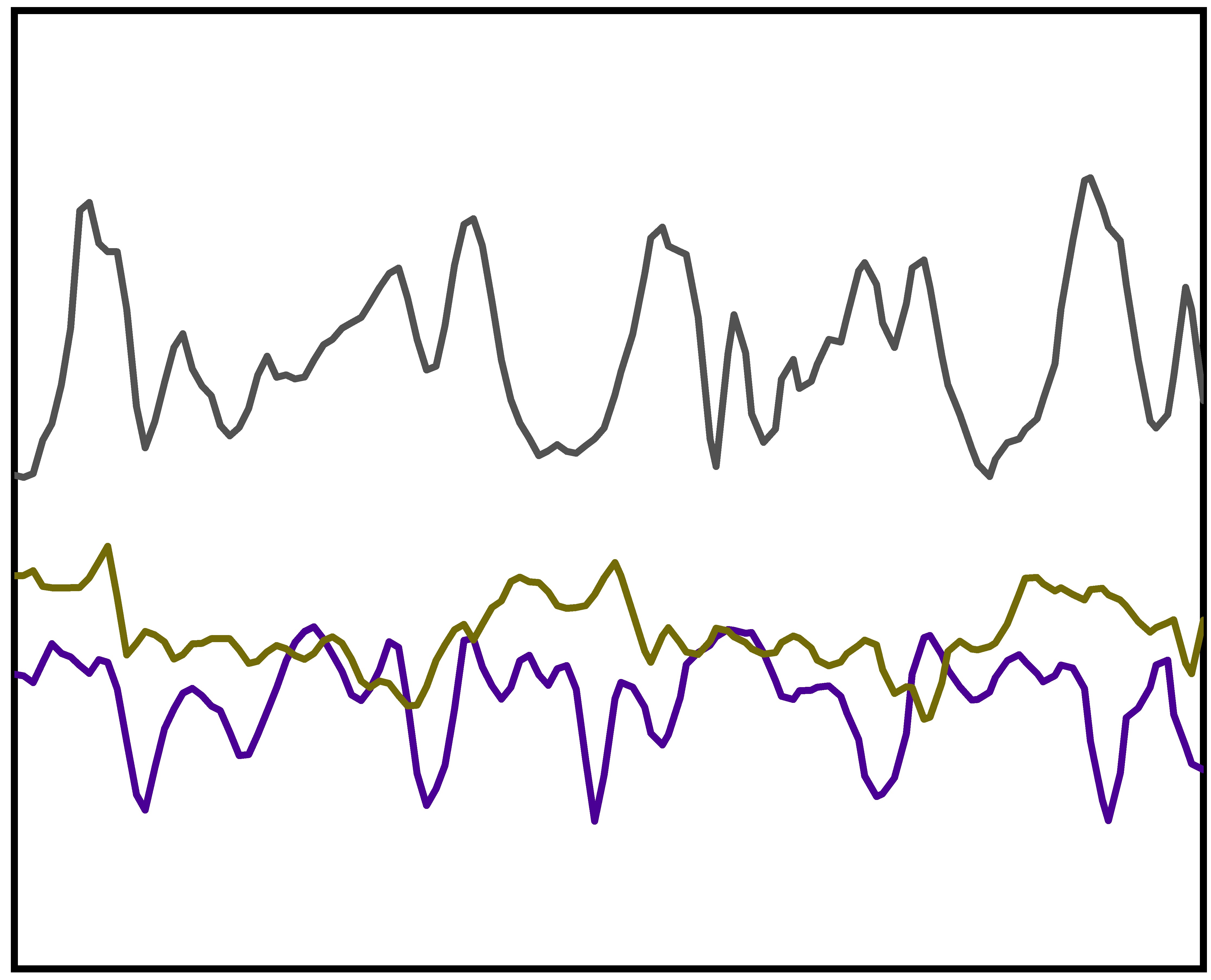}%
    \label{a2f}}
    \caption{\textbf{Effect of various types of augmentation of the sample data. (a) Raw sample data collected from $\mathbf{3}$ axis accelerometer, with (b) scaling, (c) jittering, (d) permutation, (e) magnitude warping, and (f) time warping applied on raw data.}}
    \label{au}
\end{figure*}

\subsection{Proposed Multi-Stage Training Scheme}
In the proposed training method, a number of training stages have been utilized to combine features from different transformed spaces. In Fig.~\ref{t2}, this scheme is represented schematically. These optimizations of individually trained feature extractors can be done in two stages or number of sequential stages. Algorithm~\ref{a1} and~\ref{a2} are executed for implementing two-stage training scheme, and multi-stage sequential training scheme, respectively. Operations performed in different stages are described below.

\begin{enumerate}
    \item \textbf{Individual training stage:} This stage is common for both two-stage and multi-stage training schemes. In this stage, separate CNN base parts with associate dense classifiers are trained individually to prepare the CNN base part as an efficient feature extractor for the respective transformed domain, as shown in Fig.~\ref{t1a}. Here, the identity transform is also used to incorporate features from unaltered raw data along with other transformations. However, some of these transformations contain more distinctive features related to the final activity recognition compared to others that lead to variations of performance after being trained. 
    \item \textbf{Combined training stage:} After the first training stage, each CNN base part provides an effective feature vector from its respective transformed domain.
    An additional combined training stage is employed to combine all these individually trained feature extractors for the proposed two-stage training scheme, as shown in Fig.~\ref{t1b}.  Though these architectures are similar in structure, for being trained with different representations of the transformed time series data, their extracted features will contain diverse characteristics. In this stage, all individual top dense classifier blocks trained in the first stage are removed while all CNN base parts are used unaltered as they are finely tuned as efficient feature extractors. Next, a separate densely connected layer is introduced on top of each CNN base part to reduce the extracted spatial/temporal features into more general representation. These separate densely connected layers act as the weighting vectors for feature selection from different transformed domains as introduced in Fig.~\ref{f1}. Here, the number of nodes in these densely connected layers are varied for incorporating more features from the feature extractors that contain more information for final classification. However, the information quantity of features extracted by individual CNN base parts can be analyzed by observing their performance in the first individual training stage. Following that, output feature vectors from these densely connected layers are concatenated and undergo through a combined dense classifier block. This block consisting series of densely connected layers will explore all the extracted features from different transformed domains as a whole and merge them to the final prediction with the softmax classifier at the end. In this stage, all the newly introduced densely connected layers are optimized through further training with data while keeping the CNN base part unaltered as efficient feature extractors, as shown in Fig.~\ref{t1b}.
     
     \item \textbf{Sequential training stages:} In the proposed multi-stage sequential training scheme, individually trained feature extractors are optimized and converged in a unified architecture through series of sequential training stages, as shown in Fig.~\ref{t2b}, ~\ref{t2c} and~\ref{t2d}. In this approach, two of the CNN base units operating on different transformed spaces are optimized together at a time by training an individual densely connected layer for each of the base units followed by feature concatenation and combined dense classifier unit, as shown in Fig.~\ref{t2b}. Later, these combined two feature extractors are considered as an individual unit and further merged with the next CNN base part. Similarly, in the next stage, another separate densely connected layers with a combined dense classifier unit are trained, as shown in Fig.~\ref{t2c}~and~\ref{t2d}. Therefore, through each training stage, a new CNN base part corresponding to another transformation is combined with the merged feature extractor. Moreover, each such stage merges these base feature extractor units by introducing a newly trained densely connected layers for providing the most optimized features at a whole utilizing all the existing features. As this approach optimizes two architectures at a time and contributes the merged architecture to the next training stage with the separate densely connected layers discarding the classifier unit, it provides more opportunity to empirically select the number of nodes of the separate dense layers used for feature selection and concatenation. Moreover, the number of parameters to be trained in a single stage is also lower compared to the previous combined training stage and thus provides more opportunity to extract more general features combining all these feature extractors in the expense of an increased number of training stages. Additionally, this sequential training approach is highly scalable that can incorporate a large number of feature spaces. Hence, features from additional space can be easily integrated into the feature extraction process by utilizing additional training stages with separate feature extractors. 
\end{enumerate}

\subsection{Data Augmentation}
As imbalance in the dataset makes the training process complicated for learning the distribution of minority class, data augmentation is a viable approach to mitigate such problems. In this work, we have utilized the combination of five techniques that incorporate realistic variations in the data and make the process more robust~\cite{aug}. However, all such augmentations are applied to the training data leaving the testing data unaltered for proper evaluation of the proposed methods. Jittering simulates the randomness of additive thermal noise and environmental perturbations to the acquired sensor data while scaling simulates the effect of changing the sensor's dynamic range. Moreover, in permutation operation, the input time window is divided into several segments and these segments are randomly permuted to create a modified window to make the trainer robust against the change in the sequence of steps on a particular activity. In magnitude warping, a smoothly varying random noise is multiplied with the original time series signal to warp the magnitude to simulate some random multiplicative noises that can be present in the real scenario, while in time warping, the sampling interval is smoothly varied to introduce variations in the time window. In Fig.~\ref{au}, the individual effect of these augmentations are shown on raw sample data. To increase the diversity of the augmentation process, we have used all five augmentation techniques sequentially to generate each augmented sample. Hence, in each sample, the effects of all five techniques are present that provide more realistic random variations in the augmented samples. As there exists an imbalance in the number of samples per-class in all three databases used in this study, the proposed augmentation process is applied in a higher rate to the minority classes for generating more number of augmented samples to balance out the training samples per activity class. Hence, a higher number of synthetic samples are generated for the classes with a smaller number of samples.    

\section{Results and Discussions}

Three publicly available datasets used for this study are described below. Detailed comparative analysis of the results obtained is discussed later.
\subsection{Description of the Database}
UCI HAR database~\cite{m19} contains $6$ activities collected from $30$ subjects with $50$ Hz sampling rate using $3$ axis accelerometer, gyroscope, and magnetometer embedded on a smartphone placed on the waist. USC HAR database~\cite{m20} contains $12$ activities collected from $14$ subjects with $100$ Hz sampling rate using $3$ axis accelerometer and gyroscope. SKODA database~\cite{m23} contains $11$ activities collected from a single subject in a car maintenance scenario using only a $3$ axis accelerometer sampled at $98$ Hz.

\begin{table}[!t]
\scriptsize
\centering
\caption{\textbf{Confusion Matrix on UCI HAR Dataset \cite{m19} for Proposed Two-stage Training on a Test Fold}}
\label{t3}
\begin{tabular}{|c|c|c|c|c|c|c|}
\hline
\multirow{2}{*}{\textbf{Actual}} & \multicolumn{6}{c|}{\textbf{Predicted}}                                                                                                                                                      \\ \cline{2-7} 
                                 & \textbf{Walk} & \textbf{\begin{tabular}[c]{@{}c@{}}Up\\ stairs\end{tabular}} & \textbf{\begin{tabular}[c]{@{}c@{}}Down\\ stairs\end{tabular}} & \textbf{Sit} & \textbf{Stand} & \textbf{Lay} \\ \hline
\textbf{Walk}                    & \textbf{466}  & 20                                                           & 5                                                              & 0            & 0              & 0            \\ \hline
\textbf{Upstairs}                & 7             & \textbf{525}                                                 & 0                                                              & 0            & 0              & 0            \\ \hline
\textbf{Downstairs}              & 0             & 0                                                            & \textbf{537}                                                   & 0            & 0              & 0            \\ \hline
\textbf{Sit}                     & 0             & 0                                                            & 0                                                              & \textbf{469} & 2              & 0            \\ \hline
\textbf{Stand}                   & 0             & 0                                                            & 0                                                              & 6            & \textbf{414}   & 0            \\ \hline
\textbf{Lay}                     & 0             & 0                                                            & 0                                                              & 0            & 0              & \textbf{495} \\ \hline
\end{tabular}
\label{c1}
\end{table}
\begin{table}[!t]
\scriptsize
\centering
\caption{\textbf{Confusion Matrix on UCI HAR Dataset \cite{m19} for Proposed Multi-stage Training on a Test Fold}}
\label{t3}
\begin{tabular}{|c|c|c|c|c|c|c|}
\hline
                                  & \multicolumn{6}{c|}{\textbf{Predicted}}                                                                                                                                                                                                                                                   \\ \cline{2-7} 
\multirow{-2}{*}{\textbf{Actual}} & \textbf{Walk}                        & \textbf{\begin{tabular}[c]{@{}c@{}}Up\\ stairs\end{tabular}} & \textbf{\begin{tabular}[c]{@{}c@{}}Down\\ stairs\end{tabular}} & \textbf{Sit}                         & \textbf{Stand}                       & \textbf{Lay}                         \\ \hline
\textbf{Walk}                     & \cellcolor[HTML]{FFFFFF}\textbf{478} & 9                                                            & 4                                                              & 0                                    & 0                                    & 0                                    \\ \hline
\textbf{Upstairs}                 & 4                                    & \cellcolor[HTML]{FFFFFF}\textbf{528}                         & 0                                                              & 0                                    & 0                                    & 0                                    \\ \hline
\textbf{Downstairs}               & 0                                    & 0                                                            & \cellcolor[HTML]{FFFFFF}\textbf{537}                           & 0                                    & 0                                    & 0                                    \\ \hline
\textbf{Sit}                      & 0                                    & 0                                                            & 0                                                              & \cellcolor[HTML]{FFFFFF}\textbf{470} & 1                                    & 0                                    \\ \hline
\textbf{Stand}                    & 0                                    & 0                                                            & 0                                                              & 3                                    & \cellcolor[HTML]{FFFFFF}\textbf{417} & 0                                    \\ \hline
\textbf{Lay}                      & 0                                    & 0                                                            & 0                                                              & 0                                    & 0                                    & \cellcolor[HTML]{FFFFFF}\textbf{495} \\ \hline
\end{tabular}
\label{c2}
\end{table}

\begin{table}[!t]
\scriptsize
\centering
\caption{\textbf{Average Cross-Validation Performance Analysis on Various Activities of UCI HAR Dataset \cite{m19} for Proposed Two-Stage and Multi-Stage Training}}
\label{c3}
\begin{tabular}{|c|c|c|c|c|c|c|c|}
\hline
\multirow{3}{*}{\textbf{\begin{tabular}[c]{@{}c@{}}Met-\\ rics\end{tabular}}}    & \multirow{3}{*}{\textbf{\begin{tabular}[c]{@{}c@{}}Prop.\\ Meth.\end{tabular}}} & \multicolumn{6}{c|}{\textbf{Class}}                                                                                                                                                                                                                                                                \\ \cline{3-8} 
                                                                                 &                                                                                 & \multirow{2}{*}{\textbf{Walk}} & \multirow{2}{*}{\textbf{\begin{tabular}[c]{@{}c@{}}Up\\ Stairs\end{tabular}}} & \multirow{2}{*}{\textbf{\begin{tabular}[c]{@{}c@{}}Down\\ Stairs\end{tabular}}} & \multirow{2}{*}{\textbf{Sit}} & \multirow{2}{*}{\textbf{Stand}} & \multirow{2}{*}{\textbf{Lay}} \\
                                                                                 &                                                                                 &                                &                                                                               &                                                                                 &                               &                                 &                               \\ \hline
\multirow{2}{*}{\textbf{\begin{tabular}[c]{@{}c@{}}Prec.\\ (\%)\end{tabular}}}   & \textbf{2-Stg.}                                                                 & 98.53                          & 96.34                                                                         & 99.14                                                                           & 98.75                         & 99.58                           & 100                           \\ \cline{2-8} 
                                                                                 & \textbf{M-Stg.}                                                                 & 99.27                          & 98.36                                                                         & 99.32                                                                           & 99.46                         & 99.81                           & 100                           \\ \hline
\multirow{2}{*}{\textbf{\begin{tabular}[c]{@{}c@{}}Rec.\\ (\%)\end{tabular}}}    & \textbf{2-Stg.}                                                                 & 94.93                          & 98.72                                                                         & 100                                                                             & 99.61                         & 98.64                           & 100                           \\ \cline{2-8} 
                                                                                 & \textbf{M-Stg.}                                                                 & 97.44                          & 99.26                                                                         & 100                                                                             & 99.83                         & 99.35                           & 100                           \\ \hline
\multirow{2}{*}{\textbf{\begin{tabular}[c]{@{}c@{}}IoU Sc.\\ (\%)\end{tabular}}} & \textbf{2-Stg.}                                                                 & \multicolumn{1}{l|}{96.31}     & \multicolumn{1}{l|}{97.41}                                                    & \multicolumn{1}{l|}{99.49}                                                      & \multicolumn{1}{l|}{99.07}    & \multicolumn{1}{l|}{99.02}      & \multicolumn{1}{l|}{100}      \\ \cline{2-8} 
                                                                                 & \textbf{M-Stg.}                                                                 & \multicolumn{1}{l|}{98.24}     & \multicolumn{1}{l|}{98.68}                                                    & \multicolumn{1}{l|}{99.52}                                                      & \multicolumn{1}{l|}{99.59}    & \multicolumn{1}{l|}{99.47}      & \multicolumn{1}{l|}{100}      \\ \hline
\end{tabular}
\end{table}

\begin{figure}[!t]
    \centering
    \includegraphics[scale=0.38]{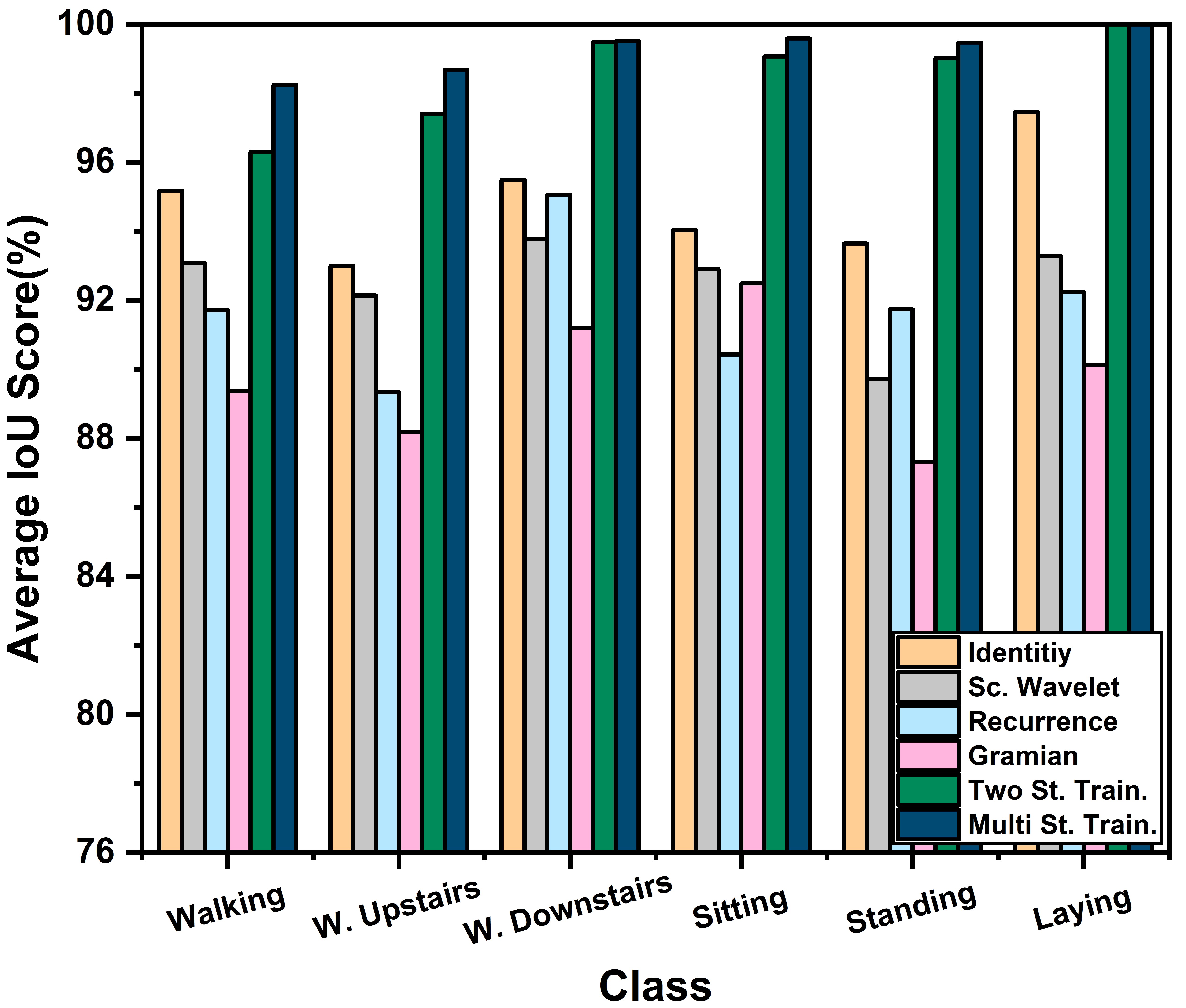}
    \caption{\textbf{Comparison of Average Cross-Validation IoU scores on various activities of UCI HAR Database~\cite{m19} obtained using different transformation schemes along with the proposed combined schemes of two-stage and multi-stage training.} }
    \label{p1}
\end{figure}

\subsection{Experimentation}

A five-fold cross-validation scheme is carried out for evaluation of the proposed scheme on each database separately. The performances of the evaluation metrics obtained from each test fold are averaged to get the final values.  All the augmentation techniques were applied to training data only. Adam optimizer (learning rate = $0.0001$, $\beta1$ = $0.9$ and $\beta2$ = $0.99$) was employed for optimization with categorical cross-entropy as loss function ($\mathscr{L}$). Keras deep learning library was used with python programming language for the implementation of the proposed neural networks. The Wilcoxon rank-sum test is used for statistical analysis of the average accuracy improvement obtained from the proposed scheme. The accuracies of the proposed schemes are statistically analyzed and the statistical significance level is set to $\alpha = 0.01$. The null hypothesis is that no significant improvement of average accuracy is achieved using the proposed scheme over the other existing best performing approaches. 

\subsection{Performance Evaluation}
The performance of the optimized networks is evaluated using the test data of various datasets. Traditional evaluation metrics for the multi-class classification task, i.e accuracy, precision, recall, and intersection-over-union (IoU) score, are employed for analyzing the performance. In Tab.~\ref{c1} and~\ref{c2}, confusion matrices are provided for the proposed two-stage and multi-stage training approach on UCI HAR database~\cite{m19} on a specific test fold.  Moreover, in Tab.~\ref{c3}, the score of averaged cross-validation evaluation metrics are provided for both these training approaches.  It is clear that both these approaches provide a considerable performance of over $98\%$ in most of these classes that are separated almost perfectly. However, the two-stage method slightly struggles to separate features between walking and ascending upstairs activities as these activities contain close inter-relation in the feature space. But, in the case of multi stage-training, this problem is reduced which signifies the robust optimization capability of this method as it can separate features with proximity.

In Fig.~\ref{p1}, the average cross-validation IoU score of the optimized networks on different transformed spaces along with the final converged networks using both two-stage and multi-stage training are compared for all the activities. It is visible that identity transform representing the unaltered raw data provides better performance with more than $2\%$ improvement in most classes compared to other transformed spaces in case of individual training. However, irrespective of the performance, all the networks operating on separate transformed spaces extract features that are significantly different as they work with diversified representations of the data. Through optimization of these features, as visible in Fig.~\ref{p1}, the proposed two-stage, and multi-stage training approach provide a sharp increase in IoU scores in all the activity classes compared to the individual training stage. However, lower performing transformed spaces are de-emphasized through a smaller number of densely connected nodes and with smaller weights generated in the later training stages while merging, as shown in Fig.~\ref{t2}. For example, in two-stage training configuration (Fig.~\ref{t1b}) for UCI database, before concatenation of features extracted from multiple transformed spaces, 96 densely connected nodes are provided following the identity transformed feature space while 32 nodes are provided following the GAF transformed space as identity transformed features provided $5.01\%$ higher average IoU score compared to the GAF transformed space in the individual training stage. Hence, more number of nodes can be adjusted for emphasizing the individually better-performing feature space. Despite that, all of the transformed spaces contribute some new and valuable information that may be indistinguishable even on other space that provides significantly better performance. Moreover, the later training stages are mainly dedicated to extracting the most distinguishable features while de-emphasizing the redundant features and thus provides this higher IoU score.  

Moreover, in multi-stage sequential training, two of the feature spaces are optimized at a time by integrating an additional feature space to the resultant feature space (Fig.~4(c)-4(e)). It should be noticed that more number of nodes are provided in the densely connected layer following the features space of respective transformation to emphasize the features from those space that provided higher performance during the individual training stage. For example, in Fig 4(c), 144 nodes are provided for the identity transformed feature space while 112 nodes for the scattering wavelet transformed space, as features from identity transformed space performed better in the individual training stage. This manipulation of the number of nodes is iteratively done to reach maximum performance. If the sequence of optimization is altered, we have to adjust the number of nodes that were applied to the extracted features of different transformed spaces. It is expected that in any combination, the achieved performance will be similar if the number of nodes for different transformed spaces is properly adjusted which ensures the proper exploration of the generated feature spaces. However, in the sequential integration process, we have integrated individually better-performing feature spaces in early stages with more number of nodes before feature concatenation.

\begin{table}[!t]
\centering
\caption{\textbf{Number of Trainable and Non-trainable Parameters on Various Training Stages for Proposed Training Schemes in UCI HAR Database~\cite{m19}}}
\label{tr}
\begin{tabular}{|c|c|c|c|c|}
\hline
\multirow{2}{*}{\textbf{\begin{tabular}[c]{@{}c@{}}Training\\ Stage\end{tabular}}} & \multicolumn{2}{c|}{\textbf{Two Stage Training}}                                       & \multicolumn{2}{c|}{\textbf{Multi Stage Training}}                                     \\ \cline{2-5} 
                                                                                   & \textbf{Trainable} & \textbf{\begin{tabular}[c]{@{}c@{}}Non-\\ Trainable\end{tabular}} & \textbf{Trainable} & \textbf{\begin{tabular}[c]{@{}c@{}}Non-\\ Trainable\end{tabular}} \\ \hline
\multirow{2}{*}{\textbf{Stage-1}}                                                  & 1.2M (1D)          & \multirow{2}{*}{0}                                                & 1.2M (1D)          & \multirow{2}{*}{0}                                                \\ \cline{2-2} \cline{4-4}
                                                                                   & 2.8M (2D)          &                                                                   & 2.8M (2D)          &                                                                   \\ \hline
\textbf{Stage-2}                                                                   & 82K                & 8.2M                                                              & 82K                & 2.5M                                                              \\ \hline
\textbf{Stage-3}                                                                   & -                  & -                                                                 & 82K                & 5.4M                                                              \\ \hline
\textbf{Stage-4}                                                                   & -                  & -                                                                 & 82K                & 8.3M                                                              \\ \hline
\end{tabular}
\end{table}

\begin{table*}[!t]
\scriptsize
\centering
\caption{\textbf{Confusion Matrix on USC HAR Dataset \cite{m20} for Proposed Two-stage Training on a Test Fold}}
\label{usc1}
\begin{tabular}{|c|c|c|c|c|c|c|c|c|c|c|c|}
\hline
                                  & \multicolumn{11}{c|}{\textbf{Predicted}}                                                                                                                                                                                                                                                                                                                                                                                                                                                                                                                                                                       \\ \cline{2-12} 
\multirow{-2}{*}{\textbf{Actual}} & \textbf{\begin{tabular}[c]{@{}c@{}}Walking\\ Forward\end{tabular}} & \textbf{\begin{tabular}[c]{@{}c@{}}Walk.\\ Left\end{tabular}} & \textbf{\begin{tabular}[c]{@{}c@{}}Walk.\\ Right\end{tabular}} & \textbf{\begin{tabular}[c]{@{}c@{}}Walk.\\ Upst.\end{tabular}} & \textbf{\begin{tabular}[c]{@{}c@{}}Walk.\\ Downst.\end{tabular}} & \textbf{Running}                     & \textbf{Jumping}                     & \textbf{Sitting}                      & \textbf{Standing}                    & \textbf{Sleeping}                     & \textbf{\begin{tabular}[c]{@{}c@{}}In\\  Elevator\end{tabular}} \\ \hline
\textbf{W. Forward}               & \cellcolor[HTML]{FFFFFF}\textbf{1584}                              & 4                                                             & 8                                                              & 7                                                              & 1                                                                & 1                                    & 0                                    & 0                                     & 2                                    & 0                                     & 0                                                               \\ \hline
\textbf{W. Left}                  & 8                                                                  & \cellcolor[HTML]{FFFFFF}\textbf{1066}                         & 5                                                              & 2                                                              & 3                                                                & 1                                    & 0                                    & 0                                     & 0                                    & 0                                     & 0                                                               \\ \hline
\textbf{W. Right}                 & 5                                                                  & 3                                                             & \cellcolor[HTML]{FFFFFF}\textbf{1101}                          & 1                                                              & 2                                                                & 0                                    & 1                                    & 0                                     & 1                                    & 0                                     & 0                                                               \\ \hline
\textbf{W. Upstairs}              & 1                                                                  & 2                                                             & 5                                                              & \cellcolor[HTML]{FFFFFF}\textbf{893}                           & 2                                                                & 1                                    & 2                                    & 0                                     & 1                                    & 0                                     & 0                                                               \\ \hline
\textbf{W. Down.}                 & 2                                                                  & 3                                                             & 4                                                              & 1                                                              & \cellcolor[HTML]{FFFFFF}\textbf{846}                             & 2                                    & 4                                    & 0                                     & 0                                    & 0                                     & 0                                                               \\ \hline
\textbf{Running}                  & 2                                                                  & 3                                                             & 3                                                              & 0                                                              & 3                                                                & \cellcolor[HTML]{FFFFFF}\textbf{711} & 2                                    & 0                                     & 2                                    & 0                                     & 0                                                               \\ \hline
\textbf{Jumping}                  & 2                                                                  & 1                                                             & 0                                                              & 2                                                              & 3                                                                & 0                                    & \cellcolor[HTML]{FFFFFF}\textbf{416} & 1                                     & 2                                    & 0                                     & 1                                                               \\ \hline
\textbf{Sitting}                  & 0                                                                  & 0                                                             & 0                                                              & 0                                                              & 0                                                                & \cellcolor[HTML]{FFFFFF}0            & 1                                    & \cellcolor[HTML]{FFFFFF}\textbf{1015} & 6                                    & 0                                     & 2                                                               \\ \hline
\textbf{Standing}                 & 1                                                                  & 0                                                             & 0                                                              & 0                                                              & 0                                                                & \cellcolor[HTML]{FFFFFF}0            & 1                                    & 8                                     & \cellcolor[HTML]{FFFFFF}\textbf{966} & 0                                     & 12                                                              \\ \hline
\textbf{Sleeping}                 & 0                                                                  & 0                                                             & 0                                                              & 0                                                              & 0                                                                & 0                                    & 3                                    & 5                                     & 0                                    & \cellcolor[HTML]{FFFFFF}\textbf{1580} & 0                                                               \\ \hline
\textbf{In Elevator}              & 0                                                                  & 0                                                             & 0                                                              & 0                                                              & 0                                                                & \cellcolor[HTML]{FFFFFF}0            & 0                                    & 0                                     & 14                                   & 1                                     & \cellcolor[HTML]{FFFFFF}\textbf{655}                            \\ \hline
\end{tabular}
\end{table*}

\begin{table*}[!t]
\scriptsize
\centering
\caption{\textbf{Confusion Matrix on USC HAR Dataset \cite{m20} for Proposed Multi-stage Training on a Test Fold}}
\label{usc2}
\begin{tabular}{|c|c|c|c|c|c|c|c|c|c|c|c|}
\hline
                                  & \multicolumn{11}{c|}{\textbf{Predicted}}                                                                                                                                                                                                                                                                                                                                                                                                                                                                                                                                                                                 \\ \cline{2-12} 
\multirow{-2}{*}{\textbf{Actual}} & \textbf{\begin{tabular}[c]{@{}c@{}}Walking\\ Forward\end{tabular}} & \textbf{\begin{tabular}[c]{@{}c@{}}Walk.\\ Left\end{tabular}} & \textbf{\begin{tabular}[c]{@{}c@{}}Walk.\\ Right\end{tabular}} & \textbf{\begin{tabular}[c]{@{}c@{}}Walk\\ Upst.\end{tabular}} & \textbf{\begin{tabular}[c]{@{}c@{}}Walk.\\ Downst.\end{tabular}} & \textbf{Running}                     & \textbf{Jumping}                     & \textbf{Sitting}                                 & \textbf{Standing}                    & \textbf{Sleeping}                     & \textbf{\begin{tabular}[c]{@{}c@{}}In\\  Elevator\end{tabular}} \\ \hline
\textbf{W. Forward}               & \cellcolor[HTML]{FFFFFF}\textbf{1590}                              & 3                                                             & 7                                                              & 5                                                             & 1                                                                & 1                                    & 0                                    & \cellcolor[HTML]{FFFFFF}{\color[HTML]{330001} 0} & 0                                    & 0                                     & 0                                                               \\ \hline
\textbf{W. Left}                  & 4                                                                  & \cellcolor[HTML]{FFFFFF}\textbf{1075}                         & 2                                                              & 1                                                             & 2                                                                & 1                                    & 0                                    & \cellcolor[HTML]{FFFFFF}0                        & 0                                    & 0                                     & 0                                                               \\ \hline
\textbf{W. Right}                 & 4                                                                  & 2                                                             & \cellcolor[HTML]{FFFFFF}\textbf{1106}                          & 2                                                             & 0                                                                & 0                                    & 0                                    & \cellcolor[HTML]{FFFFFF}0                        & 0                                    & 0                                     & 0                                                               \\ \hline
\textbf{W. Upstairs}              & 1                                                                  & 2                                                             & 3                                                              & \cellcolor[HTML]{FFFFFF}\textbf{897}                          & 1                                                                & 2                                    & 1                                    & \cellcolor[HTML]{FFFFFF}0                        & 0                                    & 0                                     & 0                                                               \\ \hline
\textbf{W. Down.}                 & 3                                                                  & 2                                                             & 3                                                              & 1                                                             & \cellcolor[HTML]{FFFFFF}\textbf{847}                             & 3                                    & 2                                    & \cellcolor[HTML]{FFFFFF}0                        & 0                                    & 0                                     & 1                                                               \\ \hline
\textbf{Running}                  & 3                                                                  & 2                                                             & 1                                                              & 1                                                             & 2                                                                & \cellcolor[HTML]{FFFFFF}\textbf{713} & 3                                    & \cellcolor[HTML]{FFFFFF}0                        & 1                                    & 0                                     & 0                                                               \\ \hline
\textbf{Jumping}                  & 1                                                                  & 2                                                             & 0                                                              & 2                                                             & 2                                                                & 0                                    & \cellcolor[HTML]{FFFFFF}\textbf{420} & 0                                                & \cellcolor[HTML]{FFFFFF}0            & 0                                     & 1                                                               \\ \hline
\textbf{Sitting}                  & 0                                                                  & 0                                                             & 0                                                              & 0                                                             & 0                                                                & \cellcolor[HTML]{FFFFFF}0            & 1                                    & \cellcolor[HTML]{FFFFFF}\textbf{1019}            & \cellcolor[HTML]{FFFFFF}4            & 0                                     & 0                                                               \\ \hline
\textbf{Standing}                 & 0                                                                  & 0                                                             & 0                                                              & 0                                                             & 0                                                                & \cellcolor[HTML]{FFFFFF}0            & 2                                    & 5                                                & \cellcolor[HTML]{FFFFFF}\textbf{972} & 0                                     & 8                                                               \\ \hline
\textbf{Sleeping}                 & 0                                                                  & 0                                                             & 0                                                              & 0                                                             & 0                                                                & 0                                    & 2                                    & 4                                                & 0                                    & \cellcolor[HTML]{FFFFFF}\textbf{1582} & 0                                                               \\ \hline
\textbf{In Elevator}              & 0                                                                  & 0                                                             & 0                                                              & 0                                                             & 0                                                                & \cellcolor[HTML]{FFFFFF}0            & 0                                    & 2                                                & 10                                   & 1                                     & \cellcolor[HTML]{FFFFFF}\textbf{657}                            \\ \hline
\end{tabular}
\end{table*}

\begin{table}[!t]
\scriptsize
\centering
\caption{\textbf{Average Cross-Validation Performance Analysis on Various Activities of USC HAR Dataset~\cite{m20} for Two-Stage and Multi-Stage Training}}
\label{usc}
\begin{tabular}{|c|c|c|c|c|c|c|}
\hline
\multirow{3}{*}{\textbf{Class}} & \multicolumn{3}{c|}{\textbf{\begin{tabular}[c]{@{}c@{}}Two stage\\ Training\end{tabular}}}                                                                                                                                                    & \multicolumn{3}{c|}{\textbf{\begin{tabular}[c]{@{}c@{}}Multi Stage\\ Training\end{tabular}}}                                                                                                                                                  \\ \cline{2-7} 
                                & \multirow{2}{*}{\textbf{\begin{tabular}[c]{@{}c@{}}Prec.\\ (\%)\end{tabular}}} & \multirow{2}{*}{\textbf{\begin{tabular}[c]{@{}c@{}}Rec.\\ (\%)\end{tabular}}} & \multirow{2}{*}{\textbf{\begin{tabular}[c]{@{}c@{}}IoU\\ (\%)\end{tabular}}} & \multirow{2}{*}{\textbf{\begin{tabular}[c]{@{}c@{}}Prec.\\ (\%)\end{tabular}}} & \multirow{2}{*}{\textbf{\begin{tabular}[c]{@{}c@{}}Rec.\\ (\%)\end{tabular}}} & \multirow{2}{*}{\textbf{\begin{tabular}[c]{@{}c@{}}IoU\\ (\%)\end{tabular}}} \\
                                &                                                                                &                                                                               &                                                                              &                                                                                &                                                                               &                                                                              \\ \hline
\textbf{Walking Forward}        & 99.2                                                                           & 98.3                                                                          & 98.5                                                                         & 99.7                                                                           & 99.4                                                                          & 99.4                                                                         \\ \hline
\textbf{Walking Left}           & 99.1                                                                           & 98.5                                                                          & 98.7                                                                         & 99.6                                                                           & 99.3                                                                          & 99.3                                                                         \\ \hline
\textbf{Walking Right}          & 99.2                                                                           & 99.4                                                                          & 99.1                                                                         & 99.5                                                                           & 99.6                                                                          & 99.5                                                                         \\ \hline
\textbf{Walking Upstairs}       & 99.3                                                                           & 98.6                                                                          & 98.8                                                                         & 99.5                                                                           & 99.1                                                                          & 99.2                                                                         \\ \hline
\textbf{Walking Down}           & 98.2                                                                           & 98.4                                                                          & 98.1                                                                         & 99.1                                                                           & 98.8                                                                          & 98.7                                                                         \\ \hline
\textbf{Running}                & 99.0                                                                           & 98.2                                                                          & 98.4                                                                         & 99.3                                                                           & 98.7                                                                          & 98.9                                                                         \\ \hline
\textbf{Jumping}                & 97.2                                                                           & 97.4                                                                          & 97.1                                                                         & 97.9                                                                           & 98.6                                                                          & 98.1                                                                         \\ \hline
\textbf{Sitting}                & 99.1                                                                           & 99.2                                                                          & 99.0                                                                         & 99.4                                                                           & 99.5                                                                          & 99.2                                                                         \\ \hline
\textbf{Standing}               & 97.5                                                                           & 98.1                                                                          & 97.8                                                                         & 98.5                                                                           & 98.8                                                                          & 98.4                                                                         \\ \hline
\textbf{Sleeping}               & 100                                                                            & 99.5                                                                          & 99.6                                                                         & 100                                                                            & 99.7                                                                          & 99.7                                                                         \\ \hline
\textbf{In Elevator}            & 98.1                                                                           & 98.3                                                                          & 98.1                                                                         & 98.4                                                                           & 98.6                                                                          & 98.4                                                                         \\ \hline
\end{tabular}
\end{table}

\begin{table}[!t]
\scriptsize
\centering
\caption{\textbf{Average Cross-Validation Performance Analysis on Various Activities of SKODA Dataset \cite{m23} for Two-Stage and Multi-Stage Training}}
\label{sk}
\begin{tabular}{|c|c|c|l|c|c|c|}
\hline
\multirow{3}{*}{\textbf{Class}}   & \multicolumn{3}{c|}{\textbf{\begin{tabular}[c]{@{}c@{}}Two stage\\ Training\end{tabular}}}                                                                                                                                                                         & \multicolumn{3}{c|}{\textbf{\begin{tabular}[c]{@{}c@{}}Multi Stage\\ Training\end{tabular}}}                                                                                                                                                  \\ \cline{2-7} 
                                  & \multirow{2}{*}{\textbf{\begin{tabular}[c]{@{}c@{}}Prec.\\ (\%)\end{tabular}}} & \multirow{2}{*}{\textbf{\begin{tabular}[c]{@{}c@{}}Rec.\\ (\%)\end{tabular}}} & \multicolumn{1}{c|}{\multirow{2}{*}{\textbf{\begin{tabular}[c]{@{}c@{}}IoU\\ (\%)\end{tabular}}}} & \multirow{2}{*}{\textbf{\begin{tabular}[c]{@{}c@{}}Prec.\\ (\%)\end{tabular}}} & \multirow{2}{*}{\textbf{\begin{tabular}[c]{@{}c@{}}Rec.\\ (\%)\end{tabular}}} & \multirow{2}{*}{\textbf{\begin{tabular}[c]{@{}c@{}}IoU\\ (\%)\end{tabular}}} \\
                                  &                                                                                &                                                                               & \multicolumn{1}{c|}{}                                                                             &                                                                                &                                                                               &                                                                              \\ \hline
\textbf{Null}                     & 95.3                                                                           & 96.4                                                                          & 95.8                                                                                              & 96.2                                                                           & 96.5                                                                          & 96.1                                                                         \\ \hline
\textbf{Write on notepad}         & 98.5                                                                           & 97.1                                                                          & 97.6                                                                                              & 99.1                                                                           & 98.6                                                                          & 98.6                                                                         \\ \hline
\textbf{Open hood}                & 95.2                                                                           & 94.5                                                                          & 94.7                                                                                              & 97.3                                                                           & 95.4                                                                          & 96.3                                                                         \\ \hline
\textbf{Close hood}               & 95.7                                                                           & 96.1                                                                          & 95.7                                                                                              & 96.5                                                                           & 96.9                                                                          & 96.5                                                                         \\ \hline
\textbf{Check gaps on front door} & 96.3                                                                           & 97.5                                                                          & 96.5                                                                                              & 97.8                                                                           & 99.1                                                                          & 98.3                                                                         \\ \hline
\textbf{Open left front door}     & 96.6                                                                           & 95.8                                                                          & 96.1                                                                                              & 97.5                                                                           & 95.7                                                                          & 96.4                                                                         \\ \hline
\textbf{Close left front door}    & 96.7                                                                           & 95.6                                                                          & 95.9                                                                                              & 97.2                                                                           & 95.9                                                                          & 96.3                                                                         \\ \hline
\textbf{Check trunk gaps}         & 98.1                                                                           & 98.4                                                                          & 98.1                                                                                              & 99.2                                                                           & 98.7                                                                          & 98.8                                                                         \\ \hline
\textbf{Open and close trunks}    & 97.2                                                                           & 98.1                                                                          & 97.4                                                                                              & 97.7                                                                           & 99.2                                                                          & 98.3                                                                         \\ \hline
\textbf{Check steering wheel}     & 97.4                                                                           & 98.5                                                                          & 97.8                                                                                              & 97.9                                                                           & 99.4                                                                          & 98.4                                                                         \\ \hline
\end{tabular}

\end{table}

\begin{figure}[!t]
    \begin{center}
            \includegraphics[scale=0.36]{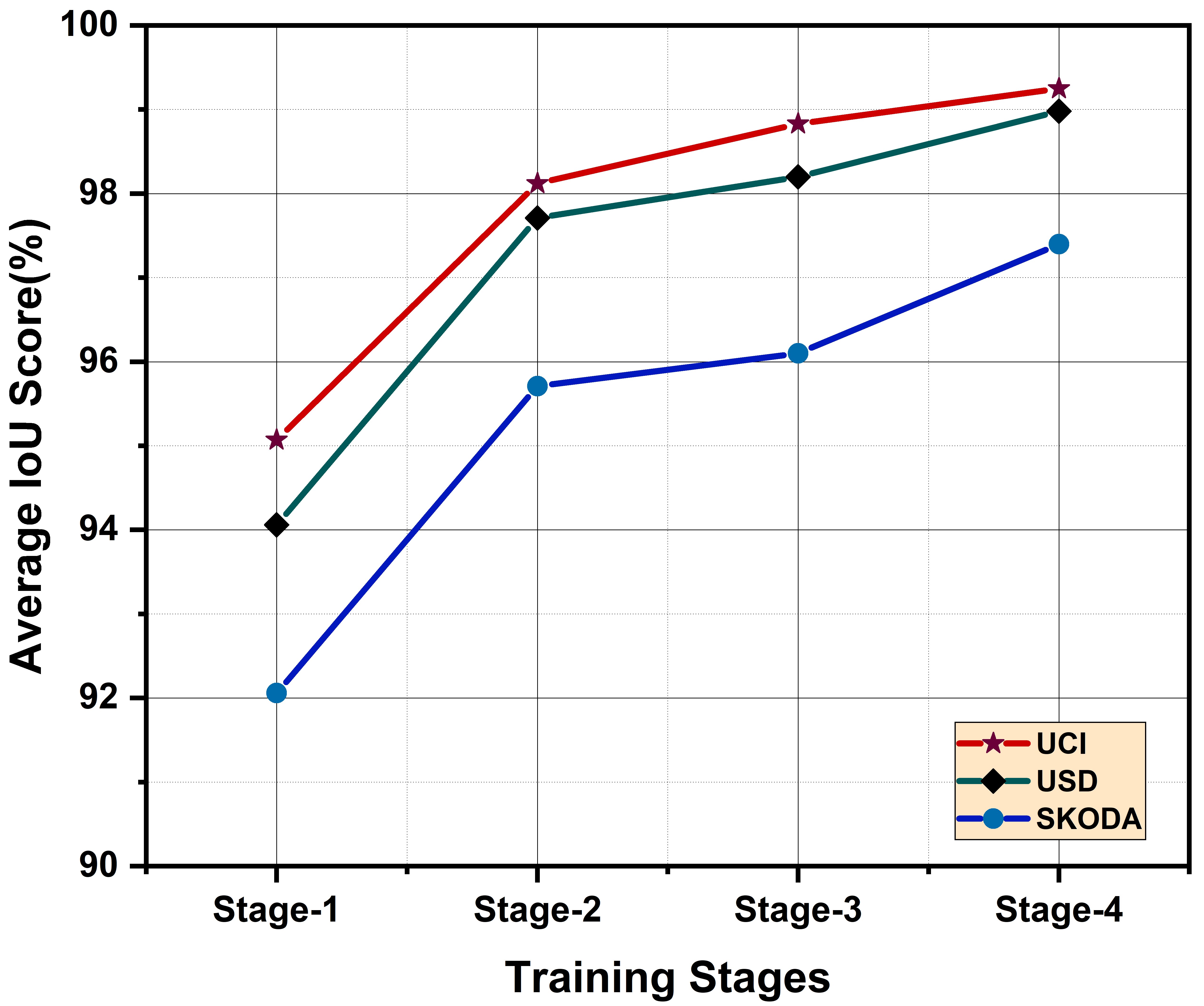}
        \caption{\textbf{Average Cross-Validation IoU score in various training stages of multi-stage sequential training on different databases.}}
        \label{seqF1}
    \end{center}
\end{figure}

In Tab.~\ref{tr}, the number of trainable and non-trainable parameters on different training stages is presented for the UCI database. As the stage-1 is similar for both these methods for training individual networks, all the parameters are set as trainable to train different feature extractors. However, 2D representation of the data requires a larger network compared to 1D representation to train with an increased size of transformed data. In later training stages, most of the parameters are set as non-trainable to utilize already trained deep feature extractors while some of the trainable parameters are introduced to merge different networks. Though the number of parameters increases in higher training stages, the number of trainable parameters on a single training stage is much smaller that makes the network resilient against overfitting with the training data. In multi-stage training, different feature extractors are merged in multiple stages and thus non-trainable parameters are increased in steps while in two-stage training, all four feature extractors are combined in one stage resulting in a larger number of non-trainable parameters in stage-2.  

\begin{table*}[!t]
\scriptsize
\centering
\caption{\textbf{Central Tendency(Mean) and Dispersion(Standard Deviation) Measures of the Average Evaluation\\ Metrics Across Various Cross-Validation Folds on Different Datasets}}
\label{ff}
\begin{tabular}{|c|c|c|c|c|c|c|c|c|}
\hline
\multirow{3}{*}{\textbf{Database}}     & \multicolumn{4}{c|}{\textbf{Two Stage training}}                                                                                                                                                                                                                                                                                                                  & \multicolumn{4}{c|}{\textbf{Multi Stage training}}                                                                                                                                                                                                                                                                                                                \\ \cline{2-9} 
                                       & \multirow{2}{*}{\textbf{\begin{tabular}[c]{@{}c@{}}Accuracy\\ (\%)\end{tabular}}} & \multirow{2}{*}{\textbf{\begin{tabular}[c]{@{}c@{}}Average\\ Precision(\%)\end{tabular}}} & \multirow{2}{*}{\textbf{\begin{tabular}[c]{@{}c@{}}Average\\ Recall(\%)\end{tabular}}} & \multirow{2}{*}{\textbf{\begin{tabular}[c]{@{}c@{}}Average\\ IoU Score(\%)\end{tabular}}} & \multirow{2}{*}{\textbf{\begin{tabular}[c]{@{}c@{}}Accuracy\\ (\%)\end{tabular}}} & \multirow{2}{*}{\textbf{\begin{tabular}[c]{@{}c@{}}Average\\ Precision(\%)\end{tabular}}} & \multirow{2}{*}{\textbf{\begin{tabular}[c]{@{}c@{}}Average\\ Recall(\%)\end{tabular}}} & \multirow{2}{*}{\textbf{\begin{tabular}[c]{@{}c@{}}Average\\ IoU Score(\%)\end{tabular}}} \\
                                       &                                                                                   &                                                                                           &                                                                                        &                                                                                          &                                                                                   &                                                                                           &                                                                                        &                                                                                          \\ \hline
UCI~\cite{m19}                                    & 98.63$\pm$0.29                                                                        & 98.75$\pm$0.27                                                                               & 98.65$\pm$0.22                                                                             & 98.55$\pm$0.23                                                                               & 99.29$\pm$0.13                                                                        & 99.37$\pm$0.08                                                                                & 99.31$\pm$0.12                                                                             & 99.25$\pm$0.11                                                                               \\ \hline
USC~\cite{m20}                                    & 98.57$\pm$0.32                                                                        & 98.71$\pm$0.21                                                                                  & 98.53$\pm$0.28                                                                             & 98.47$\pm$0.27                                                                               & 99.02$\pm$0.17                                                                        & 99.17$\pm$0.13                                                                                & 99.14$\pm$0.11                                                                             & 98.98$\pm$0.14                                                                               \\ \hline
SKODA~\cite{m23}                                  & 96.51$\pm$0.38                                                                        & 96.7$\pm$0.26                                                                                 & 96.56$\pm$0.21                                                                              & 96.72$\pm$0.29                                                                               & 97.21$\pm$0.21                                                                        & 97.64$\pm$0.18                                                                                & 97.54$\pm$0.23                                                                             & 97.41$\pm$0.14                                                                               \\ \hline
\multicolumn{1}{|c|}{\textbf{Average}} & \textbf{97.91$\pm$0.33}                                                                    & \textbf{98.05$\pm$0.25}                                                                            & \textbf{98.24$\pm$0.27}                                                                         & \textbf{97.86$\pm$0.24}                                                                           & \textbf{98.51$\pm$0.17}                                                                    & \textbf{98.73$\pm$0.13}                                                                            & \textbf{98.66$\pm$0.15}                                                                         & \textbf{98.55$\pm$0.15}                                                                           \\ \hline
\end{tabular}
\end{table*}

\begin{table*}[]
\scriptsize
\centering
\caption{\textbf{Comparison of the Proposed Schemes with Other Existing Approaches on Different Datasets}}
\label{f}
\begin{tabular}{|c|c|c|c|c|c|c|c|c|c|c|c|}
\hline
\multicolumn{4}{|c|}{\textbf{UCI HAR Database~\cite{m19}}}                                  & \multicolumn{4}{c|}{\textbf{USC HAR Database~\cite{m20}}}                   & \multicolumn{4}{c|}{\textbf{SKODA Database~\cite{m23}}}                     \\ \hline
\textbf{Work}          & \textbf{Method}        & \textbf{Acc.(\%)} & \textbf{P-value} & \textbf{Work} & \textbf{Method} & \textbf{Acc.(\%)} & \textbf{P-value} & \textbf{Work} & \textbf{Method} & \textbf{Acc.(\%)} & \textbf{P-value} \\ \hline
~\cite{r6}              & MLP         & 86.1                  & NA               & ~\cite{n4}      & MLP, J48        & 89.2                  & NA               & ~\cite{m23}     & HMM             & 86                    & NA               \\ \hline
~\cite{rr1}              & CNN                    & 94.2                  & NA               & ~\cite{m31}     & Random Forest   & 90.7                  & NA               & ~\cite{m6}      & DBN             & 89.4                  & NA               \\ \hline
~\cite{n3}              & DTW                    & 95.3                  & NA               & ~\cite{n5}      & CNN             & 93.2                  & NA               & ~\cite{n6}      & Deep Conv LSTM  & 91.2                  & NA               \\ \hline
~\cite{m19}              & SVM                    & 96                    & NA               & ~\cite{m30}     & LS-SVM          & 95.6                  & NA               & ~\cite{m11}     & CNN             & 91.7                  & NA               \\ \hline
~\cite{n1}               & Deep RNN               & 96.7                  & NA               & ~\cite{m28}     & CNN             & 97                    & NA               & ~\cite{n7}      & Ensemble LSTM   & 92.4                  & NA               \\ \hline
~\cite{n2}               & SVM                    & 97.1                  & NA               & ~\cite{n1}      & Deep RNN        & 97.8                  & NA               & ~\cite{n1}      & DeepRNN         & 92.6                  & NA               \\ \hline
\textbf{Prop. 2-Stage} & \textbf{CNN} & \textbf{98.63}   & \textbf{3.4e-5}           & \multicolumn{2}{c|}{\textbf{}}  & \textbf{98.57}    & \textbf{2.5e-6}           & \multicolumn{2}{c|}{\textbf{}}  & \textbf{96.51}    & \textbf{4.2e-4}           \\ \hline
\textbf{Prop. M-Stage} & \textbf{CNN} & \textbf{99.29}         & \textbf{5.1e-5}           & \multicolumn{2}{c|}{\textbf{}}  & \textbf{99.02}    & \textbf{1.3e-6}           & \multicolumn{2}{c|}{\textbf{}}  & \textbf{97.21}    & \textbf{2.8e-4}           \\ \hline
\end{tabular}
\end{table*}

In Tab.~\ref{usc1} and~\ref{usc2}, confusion matrices for the USC HAR dataset are provided for two-stage and multi-stage training method, respectively. Moreover,
in Tab.~\ref{usc}, the average cross-validation performance of the proposed schemes on different activity classes of this dataset is provided. It is clear that both these approaches provide consistent performance over $99\%$ for most of the classes. However, multi-stage training provides a slight increase in incorrect predictions for some closely related activities like among various walking actions, between standing and sitting actions. In Tab.~\ref{sk}, the average crosss-validation performance of both the training approaches is presented on the SKODA dataset. Though most of the activities contain close inter-relation in this dataset, our proposed training methods provide consistent performance over $95\%$ for most of the classes. However, some activities like opening and closing hood, opening, and closing doors, are difficult to separate as expected. Despite that, comparable performances have been achieved in these classes utilizing the proposed scheme.  

In Fig.~\ref{seqF1}, the average IoU score in different stages are shown for multi-stage training. It is clear that each stage provides some improvement in performance by incorporating new features. However, in the first two stages, the trained network has achieved significant performance improvement with more than $3\%$ improvement in the average IoU score mostly achieved utilizing the features from identity transformation and scattering wavelet transformation with the 1D deep CNN feature extractor. Nevertheless, features from other transformations exploited at the later stages still provide a considerable contribution with around $1\sim2\%$ improvement in total to make the final network more optimized to separate challenging classes and thus to attain a higher average IoU score. Hence, integration of features from four transformed spaces in the proposed sequential training approach, $4\sim6\%$ improvement of average IoU score is achieved in total compared to operating with raw sensor data alone. In Tab.~\ref{ff}, central tendency (mean) and dispersion measures (standard deviation) of the evaluation metrics in different databases are provided. It should be noticed that the standard deviations of performance over various cross-validation folds are trivial in most cases that signify the generalizability of the proposed scheme. Moreover, around $50\%$ reduction of standard deviation is achieved in the multi-stage training approach over the two-stage counterpart. Additionally, the average performances over all three databases are also reported which surpasses $98\%$ in most cases that signify the robustness and consistency of the proposed scheme over numerous databases.

Various existing approaches are compared with the proposed ones in Tab.~\ref{f} on different datasets. Average accuracies obtained from the proposed two-stage and multi-stage training methods are compared with the reported accuracy of varieties of state-of-the-art approaches. It can be noted that the proposed multi-stage scheme has improved average accuracy from 86.1\% to 99.29\% (13.19\% improvement) in UCI database, from 89.2\% to 99.02\%  (9.82\% improvement) in USC database, and from 86\% to 97.21\% (11.21\% improvement) in SKODA database. The improvement in the multi-stage approach is around $1\%$ higher over the two-stage training approach for its increased opportunity of optimization through multiple stages. However, the training complexity also increases as more number of training stages need to be adjusted. As the $p-$values obtained from the statistical significance test on different databases are considerably smaller from the predefined threshold of 0.01, we have to reject the null hypothesis and it suggests that considerable improvement of average accuracy is achieved using the proposed schemes over other existing approaches. Moreover, the following observations can be drawn from the analysis:

\begin{itemize}
    \item In the UCI database, the shallow machine learning approaches (\cite{n2,m19}) comparatively provide better performance compared to other traditional deep learning-based approaches (\cite{m28, rr1}) due to the smaller amount of available training data. It should be noticed that the proposed scheme exploits the available data by incorporating a diverse representation of the feature space through multiple training stages that extract the effective features without the overfitting issues which is predominant in other traditional deep learning approaches. Hence, despite the smaller amount of available data, the proposed multi-stage deep CNN-based approach outperforms traditional shallow classifiers in the UCI database.
    
    \item Deep learning-based approaches mostly dominate in the USC database due to the higher number of training samples. Nevertheless, due to more number of activity classes, there exist additional complexity in the feature extraction process. These traditional deep learning-based methods mostly struggle in challenging cases due to the complicated deep network-based approaches that operate solely on raw sensor data. On the contrary, the proposed scheme employs deep feature extractors efficiently on numerous transformed spaces and splits the training process into several stages that provides a better selection of features along with higher resilience over random noises and perturbations.
    
    \item The SKODA database is more complicated due to close inter-relation of many activities along with a large number of activity classes that result in smaller performance in most-other approaches. However, the proposed scheme explores a number of representational feature spaces instead of a single space that not only increases the diversity of features but also assists better separation of inter-related activities. Hence, the proposed scheme provides a sharp improvement of average accuracy (more than $4.6\%$ in multi-stage over the other best approach) in this database. 
    
\end{itemize}

Though we have incorporated features from four transformed spaces (including identity transform) in this work, it is to be noted that the proposed sequential training scheme is adaptive and features from newly transformed spaces can be easily integrated with the resultant feature space by including additional training stages. However, to incorporate effective features from new representational space, separate CNN-based feature extractors need to be incorporated which will increase the total size of the network accordingly. But, in the traditional training approach, if the whole system of the network is trained in a single training stage, it will be very complicated to achieve convergence and the network will be highly prone to overfit with the training data that will limit the integration of numerous transformations. Whereas, the proposed training scheme separately optimizes individual deep feature extractors and integrates the extracted feature spaces in a sequential manner that makes it possible to exploit a large number of feature spaces which provides a significant advantage over the traditional approaches.
However, if a large number of transformed spaces are integrated into the feature extraction process, the increased size of the network may limit its application in mobile devices. Nevertheless, it is shown that very satisfactory performance is achieved by incorporating a fewer number of transformed spaces only. Hence, to reduce the complexity of the network for mobile applications, a fewer number of transformed spaces can be integrated into the feature extraction process while for achieving more robust performance, a large number of transformed feature spaces can be utilized.

\section{Conclusion}
In this paper, various types of human activities are recognized utilizing the proposed multi-stage training method. Firstly, the raw data undergo through numerous transformations to interpret the information encoded in raw data in different spaces and thus to obtain a diversified representation of the features. Afterwards, separate deep CNN architectures are trained on each space to be an optimized feature extractor from that particular space for the final prediction of activity. Later, these tuned feature extractors are merged into a final form of deep network effectively through a combined training stage or through sequential stages of training by exploring the extracted feature spaces exhaustively to attain the most robust and accurate feature representation. It is found that instead of utilizing trained CNN as a feature extractor from a single space if multiple trained CNNs dealing with numerous transformed spaces can be utilized together, much better representation of features can be obtained.  Such an idea of multiple training stages utilizing the initially trained CNN models from the preceding stages operating on different transformed spaces can offer a significant increase in performance with $4\sim6\%$ improvement in average IoU scores. This method outperforms other state-of-the-art approaches in different datasets by a considerable margin with an average accuracy of $98.51\%$ (11.49\% average improvement) over three databases. Therefore, the proposed scheme opens up a new approach of employing multiple training stages for deep CNNs deploying various transformed representations of data which can also be utilized in very diversified applications by increasing the diversity of the extracted features. 
 
\bibliographystyle{IEEEtran}
\bibliography{ref}

\begin{thebibliography}{10}
\providecommand{\url}[1]{#1}
\csname url@samestyle\endcsname
\providecommand{\newblock}{\relax}
\providecommand{\bibinfo}[2]{#2}
\providecommand{\BIBentrySTDinterwordspacing}{\spaceskip=0pt\relax}
\providecommand{\BIBentryALTinterwordstretchfactor}{4}
\providecommand{\BIBentryALTinterwordspacing}{\spaceskip=\fontdimen2\font plus
\BIBentryALTinterwordstretchfactor\fontdimen3\font minus
  \fontdimen4\font\relax}
\providecommand{\BIBforeignlanguage}[2]{{%
\expandafter\ifx\csname l@#1\endcsname\relax
\typeout{** WARNING: IEEEtran.bst: No hyphenation pattern has been}%
\typeout{** loaded for the language `#1'. Using the pattern for}%
\typeout{** the default language instead.}%
\else
\language=\csname l@#1\endcsname
\fi
#2}}
\providecommand{\BIBdecl}{\relax}
\BIBdecl

\bibitem{a1}
N.~Islam, Y.~Faheem, I.~U. Din, M.~Talha, M.~Guizani, and M.~Khalil, ``A
  blockchain-based fog computing framework for activity recognition as an
  application to e-healthcare services,'' \emph{Future Generation Computer
  Systems}, vol. 100, pp. 569--578, 2019.

\bibitem{i1}
A.~Jalal, Y.-H. Kim, Y.-J. Kim, S.~Kamal, and D.~Kim, ``Robust human activity
  recognition from depth video using spatiotemporal multi-fused features,''
  \emph{Pattern recognition}, vol.~61, pp. 295--308, 2017.

\bibitem{r6}
R.-A. Voicu, C.~Dobre, L.~Bajenaru, and R.-I. Ciobanu, ``Human physical
  activity recognition using smartphone sensors,'' \emph{Sensors}, vol.~19,
  no.~3, p. 458, 2019.

\bibitem{n2}
A.~Jain and V.~Kanhangad, ``Human activity classification in smartphones using
  accelerometer and gyroscope sensors,'' \emph{IEEE Sensors Journal}, vol.~18,
  no.~3, pp. 1169--1177, 2017.

\bibitem{m29}
R.~C. Kumar, S.~S. Bharadwaj, B.~Sumukha, and K.~George, ``Human activity
  recognition in cognitive environments using sequential elm,'' in \emph{2016
  Second International Conference on Cognitive Computing and Information
  Processing (CCIP)}.\hskip 1em plus 0.5em minus 0.4em\relax IEEE, 2016, pp.
  1--6.

\bibitem{m23}
P.~Zappi, C.~Lombriser, T.~Stiefmeier, E.~Farella, D.~Roggen, L.~Benini, and
  G.~Tr{\"o}ster, ``Activity recognition from on-body sensors: accuracy-power
  trade-off by dynamic sensor selection,'' in \emph{European Conference on
  Wireless Sensor Networks}.\hskip 1em plus 0.5em minus 0.4em\relax Springer,
  2008, pp. 17--33.

\bibitem{n3}
S.~Seto, W.~Zhang, and Y.~Zhou, ``Multivariate time series classification using
  dynamic time warping template selection for human activity recognition,'' in
  \emph{2015 IEEE Symposium Series on Computational Intelligence}.\hskip 1em
  plus 0.5em minus 0.4em\relax IEEE, 2015, pp. 1399--1406.

\bibitem{m31}
P.~Vaka, F.~Shen, M.~Chandrashekar, and Y.~Lee, ``Pemar: A pervasive middleware
  for activity recognition with smart phones,'' in \emph{2015 IEEE
  International Conference on Pervasive Computing and Communication Workshops
  (PerCom Workshops)}.\hskip 1em plus 0.5em minus 0.4em\relax IEEE, 2015, pp.
  409--414.

\bibitem{m6}
M.~A. Alsheikh, A.~Selim, D.~Niyato, L.~Doyle, S.~Lin, and H.-P. Tan, ``Deep
  activity recognition models with triaxial accelerometers,'' in
  \emph{Workshops at the Thirtieth AAAI Conference on Artificial Intelligence},
  2016.

\bibitem{m30}
Y.~Zheng, ``Human activity recognition based on the hierarchical feature
  selection and classification framework,'' \emph{Journal of Electrical and
  Computer Engineering}, vol. 2015, 2015.

\bibitem{m28}
W.~Jiang and Z.~Yin, ``Human activity recognition using wearable sensors by
  deep convolutional neural networks,'' in \emph{Proceedings of the 23rd ACM
  international conference on Multimedia}, 2015, pp. 1307--1310.

\bibitem{rr1}
V.~Bianchi, M.~Bassoli, G.~Lombardo, P.~Fornacciari, M.~Mordonini, and
  I.~De~Munari, ``Iot wearable sensor and deep learning: An integrated approach
  for personalized human activity recognition in a smart home environment,''
  \emph{IEEE Internet of Things Journal}, vol.~6, no.~5, pp. 8553--8562, 2019.

\bibitem{r8}
B.~Zhou, J.~Yang, and Q.~Li, ``Smartphone-based activity recognition for indoor
  localization using a convolutional neural network,'' \emph{Sensors}, vol.~19,
  no.~3, p. 621, 2019.

\bibitem{m11}
D.~Ravi, C.~Wong, B.~Lo, and G.-Z. Yang, ``Deep learning for human activity
  recognition: A resource efficient implementation on low-power devices,'' in
  \emph{2016 IEEE 13th international conference on wearable and implantable
  body sensor networks (BSN)}.\hskip 1em plus 0.5em minus 0.4em\relax IEEE,
  2016, pp. 71--76.

\bibitem{n6}
F.~J. Ord{\'o}{\~n}ez and D.~Roggen, ``Deep convolutional and lstm recurrent
  neural networks for multimodal wearable activity recognition,''
  \emph{Sensors}, vol.~16, no.~1, p. 115, 2016.

\bibitem{n5}
Y.~Chen and Y.~Xue, ``A deep learning approach to human activity recognition
  based on single accelerometer,'' in \emph{2015 IEEE International Conference
  on Systems, Man, and Cybernetics}.\hskip 1em plus 0.5em minus 0.4em\relax
  IEEE, 2015, pp. 1488--1492.

\bibitem{n1}
A.~Murad and J.-Y. Pyun, ``Deep recurrent neural networks for human activity
  recognition,'' \emph{Sensors}, vol.~17, no.~11, p. 2556, 2017.

\bibitem{r2}
A.~Gumaei, M.~M. Hassan, A.~Alelaiwi, and H.~Alsalman, ``A hybrid deep learning
  model for human activity recognition using multimodal body sensing data,''
  \emph{IEEE Access}, vol.~7, pp. 99\,152--99\,160, 2019.

\bibitem{r9}
S.~Chung, J.~Lim, K.~J. Noh, G.~Kim, and H.~Jeong, ``Sensor data acquisition
  and multimodal sensor fusion for human activity recognition using deep
  learning,'' \emph{Sensors}, vol.~19, no.~7, p. 1716, 2019.

\bibitem{r5}
M.~Lv, W.~Xu, and T.~Chen, ``A hybrid deep convolutional and recurrent neural
  network for complex activity recognition using multimodal sensors,''
  \emph{Neurocomputing}, vol. 362, pp. 33--40, 2019.

\bibitem{r7}
H.~Yu, G.~Pan, M.~Pan, C.~Li, W.~Jia, L.~Zhang, and M.~Sun, ``A hierarchical
  deep fusion framework for egocentric activity recognition using a wearable
  hybrid sensor system,'' \emph{Sensors}, vol.~19, no.~3, p. 546, 2019.

\bibitem{g1}
Z.~Wang and T.~Oates, ``Encoding time series as images for visual inspection
  and classification using tiled convolutional neural networks,'' in
  \emph{Workshops at the Twenty-Ninth AAAI Conference on Artificial
  Intelligence}, 2015.

\bibitem{r1}
N.~Hatami, Y.~Gavet, and J.~Debayle, ``Classification of time-series images
  using deep convolutional neural networks,'' in \emph{Tenth International
  Conference on Machine Vision (ICMV 2017)}, vol. 10696.\hskip 1em plus 0.5em
  minus 0.4em\relax International Society for Optics and Photonics, 2018, p.
  106960Y.

\bibitem{w2}
S.~Mallat, ``Group invariant scattering,'' \emph{Communications on Pure and
  Applied Mathematics}, vol.~65, no.~10, pp. 1331--1398, 2012.

\bibitem{s}
W.~Lu, F.~Fan, J.~Chu, P.~Jing, and S.~Yuting, ``Wearable computing for
  internet of things: A discriminant approach for human activity recognition,''
  \emph{IEEE Internet of Things Journal}, vol.~6, no.~2, pp. 2749--2759, 2018.

\bibitem{m19}
D.~Anguita, A.~Ghio, L.~Oneto, X.~Parra, and J.~L. Reyes-Ortiz, ``A public
  domain dataset for human activity recognition using smartphones.'' in
  \emph{Esann}, 2013.

\bibitem{aug}
T.~T. Um, F.~M. Pfister, D.~Pichler, S.~Endo, M.~Lang, S.~Hirche, U.~Fietzek,
  and D.~Kuli{\'c}, ``Data augmentation of wearable sensor data for
  parkinson’s disease monitoring using convolutional neural networks,'' in
  \emph{Proceedings of the 19th ACM International Conference on Multimodal
  Interaction}, 2017, pp. 216--220.

\bibitem{m20}
M.~Zhang and A.~A. Sawchuk, ``Usc-had: a daily activity dataset for ubiquitous
  activity recognition using wearable sensors,'' in \emph{Proceedings of the
  2012 ACM Conference on Ubiquitous Computing}, 2012, pp. 1036--1043.

\bibitem{n4}
C.~Catal, S.~Tufekci, E.~Pirmit, and G.~Kocabag, ``On the use of ensemble of
  classifiers for accelerometer-based activity recognition,'' \emph{Applied
  Soft Computing}, vol.~37, pp. 1018--1022, 2015.

\bibitem{n7}
Y.~Guan and T.~Pl{\"o}tz, ``Ensembles of deep lstm learners for activity
  recognition using wearables,'' \emph{Proceedings of the ACM on Interactive,
  Mobile, Wearable and Ubiquitous Technologies}, vol.~1, no.~2, pp. 1--28,
  2017.

\end{thebibliography}

\begin{IEEEbiography}[{\includegraphics[width=1in,height=1.25in,clip,keepaspectratio]{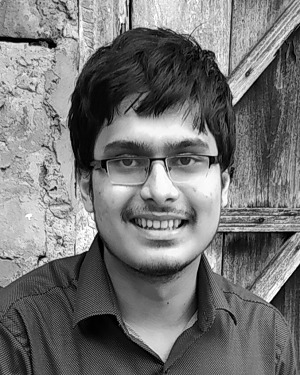}}]{Tanvir Mahmud} (S'18) received the B.Sc. degree from the EEE Department, Bangladesh University of Engineering and Technology, where he is currently pursuing the master’s degree. He is currently serving as a Lecturer with the EEE Department of Bangladesh University of Engineering and Technology. His research interest lies in VLSI circuit design, computer vision, approximate computing, biomedical signal processing, image processing, and machine learning.
\end{IEEEbiography}

\begin{IEEEbiography}[{\includegraphics[width=1in,height=1.25in,clip,keepaspectratio]{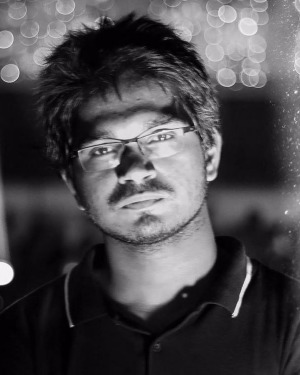}}]{A.~Q.~M.~Sazzad~Sayyed} (S'18) received the B.Sc. degree from the EEE Department, Bangladesh University of Engineering and Technology, where he is currently pursuing the master’s degree. He is currently serving as a Lecturer with the EEE Department of Southeast University, Dhaka. His research interest lies in computer vision, image processing, evolutionary computation and machine learning.
\end{IEEEbiography}

\begin{IEEEbiography}[{\includegraphics[width=1in,height=1.25in,clip,keepaspectratio]{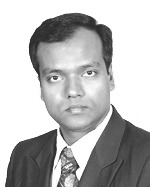}}]{Shaikh Anowarul Fattah}  (S’02–M’09–SM’16) received the B.Sc. and M.Sc. degrees from the Bangladesh University of Engineering and Technology (BUET), Bangladesh, and the Ph.D. degree in ECE from Concordia University, Canada. He held a visiting postdoctoral position and later a visiting Research Associate with Princeton University, Princeton, NJ, USA. He has been serving as a Professor with the Department of EEE, BUET. He has published over 200 inter-national journal articles and conference papers with some best paper awards. His major research interests include biomedical engineering and signal processing. He  is a Fellow of IEB. He is regularly delivering keynote/invited/visiting talks in many countries. He received several prestigious awards, such as the Concordia University’s Distinguished Doctoral Dissertation Prize in ENS, the Dr. Rashid Gold Medal (M.Sc., BUET),the NSERC Postdoctoral Fellowship, the URSI Canadian Young Scientist Award 2007, and the BAS-TWAS Young Scientists Prize 2014. He is the General Chair of IEEE R10 HTC2017, ICAICT 2020, the TPC Chair of IEEE TENSYMP 2020, IEEE WIECON-ECE 2016, 2017, MediTec 2016, IEEE ICIVPR 2017, and ICAEE 2017. He is a Committee Member of IEEEPES (LRPC), IEEE SSSIT (SDHTC), IEEE HAC (2018–2020), and R10. He was the Chair of the IEEE Bangladesh Section (2015–2016). He was the Chair of the IEEE EMBS Bangladesh Chapter (2017–2019). He is the Founder Chair of the IEEE RAS and SSIT Bangladesh Chapters. He was an Editor of the Journal of Electrical Engineering of IEB. He is an Editor of the IEEE PES eNews, an Editorial Board Member of IEEE ACCESS, and an Associate Editor of CSSP (Springer).
\end{IEEEbiography}

\begin{IEEEbiography}[{\includegraphics[width=1in,height=1.25in,clip,keepaspectratio]{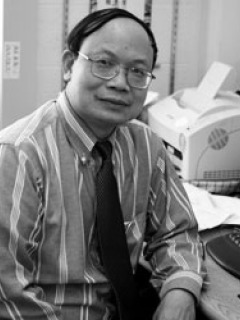}}]{Sun-Yuan Kung}  (F’88–LF’16) received the B.Eng. degree from National Taiwan University, Taipei, Taiwan, in 1971, the M.Eng. degree from the University of Rochester, Rochester, NY, USA, in 1974, and the Ph.D. degree from Stanford
University, Stanford, CA, USA, in 1977. He is currently a Professor with the Department of Electrical Engineering, Princeton University, Princeton, NJ, USA. He has authored or coauthored over 500 technical publications and numerous textbooks, including the VLSI Array Processors (Prentice-Hall, 1988), the Digital Neural Networks (Prentice-Hall, 1993), the Principal Component Neural Networks (Wiley, 1996), the Biometric Authentication: A Machine Learning Approach (Prentice-Hall, 2004), and the Kernel Methods and Machine Learning (Cambridge University Press, 2014). His current research interests include machine learning, data mining and privacy, statistical estimation, system identification, wireless communication, VLSI array processors, signal processing, and multimedia information processing. Prof. Kung was a recipient of the IEEE Signal Processing Society’s Technical Achievement Award for the contributions on parallel processing and neural network algorithms for signal processing in 1992, the IEEE Signal Processing Society’s Best Paper Award for his publication on principal component neural networks in 1996, and the IEEE Third Millennium Medal in 2000. He was a Distinguished Lecturer of the IEEE Signal Processing Society in 1994. He was a Founding Member of several technical committees of the IEEE Signal Processing Society and the first Associate Editor in VLSI area and Neural Network for the IEEE TRANSACTIONS ON SIGNAL PROCESSING in 1984 and 1991, respectively. He served as a member of the Board of Governors of the IEEE Signal Processing Society from 1989 to 1991. Since 1990, he has been the Editor-in-Chief of the Journal of VLSI Signal Processing Systems.
\end{IEEEbiography}

\end{document}